\documentclass{elsarticle}

\usepackage[utf8]{inputenc}

\usepackage{multicol}
\usepackage{xcolor}			
\usepackage{float}			
\usepackage{tabularx}		

\usepackage{enumitem}		

\usepackage{mathrsfs}
\usepackage{amsmath}
\usepackage{mathtools}	
\usepackage{amssymb}	
\usepackage{amsthm}		
\usepackage{bm}		
\usepackage{nicematrix}
\usepackage{nicefrac}
\newcommand{\IR}{{\mathbb{R}}}
\DeclarePairedDelimiter{\abs}{|}{|}

\usepackage{graphicx}
\usepackage{subcaption}
\graphicspath{{images/}}		

\usepackage{tikz}
\usepackage{tikz-3dplot}
\usetikzlibrary{arrows,angles,quotes}
\usetikzlibrary{calc}
\usetikzlibrary{shapes.geometric}

\usepackage{pgfplots}
\pgfplotsset{compat = 1.15}

\usepackage[hidelinks, pdfstartview = FitB]{hyperref} 
\usepackage[nameinlink]{cleveref}

\newcommand{\sphere}{\mathbb S}

\renewcommand{\d}[1]{\, \mathrm{d} #1}		

\colorlet{darkred}{red!60!black}
\colorlet{darkgreen}{green!40!black}
\colorlet{darkblue}{blue!60!black}

\begin{document}

\begin{frontmatter}

\title{A unified framework for grain boundary distributions in textured materials}

\author[TUBAF]{Ralf Hielscher}
\author[UH]{Rüdiger Kilian}
\author[TUBAF]{Erik Wünsche}
\author[Ox1,Ox2]{Katharina Tinka Marquardt}

\affiliation[TUBAF]{
  organization={Institute of Applied Analysis, TU Bergakademie Freiberg},
  country={Germany}
}

\affiliation[UH]{
  organization={Institute of Geoscience and Geography,   Martin-Luther-Universität Halle},
  country={German}
}

\affiliation[Ox1]{
  organization={Department of Materials, University of Oxford},
  addressline={Parks Road},
  city={Oxford OX1 3PH},
  country={United Kingdom}
}

\affiliation[Ox2]{
  organization={Zero Carbon Energy Research Institute Oxford, University of Oxford},
  addressline={Holywell House, Osney Mead},
  city={Oxford OX2 0ES},
  country={United Kingdom}
}

\begin{abstract}
  Grain boundary plane distributions are widely used to infer the mechanisms governing grain boundary formation in polycrystalline materials. We show that such interpretations are inherently ambiguous. Using a unified
eight-parameter boundary distribution framework, we derive both the grain boundary character distribution (GBCD) and the grain boundary normal distribution (GBND) and identify two limiting cases of boundary network formation.

We show that in macroscopically driven networks, the crystal-frame GBND is given by a convolution of the specimen GBND with the orientation distribution function (ODF), whereas in crystallographically driven networks the specimen GBND is
obtained by convolution of the crystal GBND with the ODF. This duality implies that anisotropy in the GBND may arise from macroscopic alignment effects rather than intrinsic crystallographic selection. Conversely, this relationship may be used to identify the dominant formation process in the measured mcirostructures.

Evaluation of a wide variety of simulated microstructures confirm the theoretically predicted relationships between texture, GBND and GBCD. In particular, our examples confirm that the GBND or GBCD alone are not sufficient for identifying grain boundary formation mechanisms.
\end{abstract}

\end{frontmatter}

\setlength{\columnsep}{2em} 

\newcommand{\sym}[2]{\item[#1] #2}

\section{Introduction}
\label{sec:introduction}
\noindent

The formation of grain boundaries in polycrystalline materials is governed by
an interplay of multiple mechanisms, including crystallographic constraints,
interface energy anisotropy, and macroscopic processes such as deformation,
growth, and recrystallization \cite{Rohrer2011,GottsteinShvindlerman2010GBMigration}. 
Depending on the dominant mechanism, grain boundary networks may exhibit preferred
crystallographic character, preferred alignment with respect to the specimen
reference frame, or a combination of both.

Over the past decades, several mathematical tools have been developed to
describe and analyze grain and phase boundary (hereafter jointly "boundary") networks in a statistical framework
\cite{AdamsField1992_GBTexture,Morawiec2004Book}. 
Central among these are the grain boundary character distribution (GBCD) and
the grain boundary normal distribution (GBND). The GBCD describes the
distribution of grain boundaries as a function of misorientation and boundary
plane orientation in crystal coordinates 
\cite{SaylorMorawiecRohrer2003_MgO5DOF,Beladi2014_Acta_GBCD_Energy,Marquardt2015,Ferreira2021}, while the GBND characterizes the distribution of boundary normals either in the specimen or crystal reference frame, \cite{GlowinskiRohrer2016_LabFrameGBNormals}. Together, these quantities provide
complementary descriptions of grain boundary networks and are widely used in the
analysis of experimental and simulated microstructures.

Despite their widespread use, the relationship between GBCD and GBND is often
interpreted without explicitly accounting for the underlying mechanisms that
govern boundary formation. In particular, it is frequently assumed that preferred
boundary plane orientations observed in the crystal reference frame directly
reflect crystallographic selection mechanisms. However, in textured materials,
preferred orientations of boundary normals may also arise from purely
macroscopic alignment effects, as we will demonstrate and quantify in this paper.

For this purpose, we propose a unified statistical framework for the analysis
of grain boundary networks based on an eight-parameter boundary distribution
function. This eight-parameter boundary distribution function was originally introduced
by Adams et~al. \cite{AdamsField1992} and comprises three parameters for each adjacent grain orientation and two parameters for the boundary normal. However, due to its high dimensionality it received little attention until now. We demonstrate in Eq.~\eqref{eq:gBND}, \eqref{eq:gbcd} that both the GBND and the GBCD can be derived directly from the eight-parameter boundary distribution.

Within our unified framework, we distinguish between two limiting cases: (i)
macroscopically driven boundary networks, in which boundary normals are
independent of the local lattice orientations, and (ii) crystallographically driven
boundary networks, in which boundary character is determined by crystallographic
mechanisms independently of the specimen reference frame.

Our key finding is that in macroscopically driven boundary networks, the GBCD can be obtained as the spherical convolution, Eq.~\eqref{eq:bnd2bndA}, of orientation distribution function (ODF) with the GBND, whereas in crystallographically driven boundary networks the GBND can be obtained as the spherical convolution, Eq.~\eqref{eq:bnda2bnd}, of the ODF with the GBCD. This duality highlights
that preferred boundary plane orientations in one reference frame may arise as a
consequence of texture rather than intrinsic crystallographic selection. 

By comparing a measured GBND with the GBND predicted from ODF and GBCD, our framework allows for the first time to quantify the extend of crystallographically driven grain boundary network formation.

For the first time, our framework enables quantification of the extent of crystallographically driven grain boundary network formation by comparing the measured GBND with the GBND predicted from the ODF and GBCD. Likewise, comparison with the GBCD allows assessment of macroscopically driven grain boundary network formation.
It further suggests that caution is
required when attributing observed boundary plane preferences to specific
mechanisms, particularly in textured materials.

Six simulated three-dimensional
microstructures covering different types of grain boundary networks (Fig.~\ref{fig:Macro}, \ref{fig:Dynamic} and \ref{fig:Mixed}) illustrate our findings. In particular, we measure the corresponding grain boundary distributions and compare them to our predictions.

It should be noted that throughout this paper we assume full knowledge of the three-dimensional microstructure. How those newly developed methods can be expanded two the study dimensional section will be subject of a forthcoming work.

All simulations, computations and analysis were carried out using Neper
\cite{QUEY20111729, quey:hal-01626440} and MTEX 6.1 \cite{MTEX}. The scripts reproducing all figures of this paper can be found at \url{https://github.com/mtex-toolbox/mtex-paper/tree/master/GBDistributions}.


\bigskip

\noindent
Throughout the paper we will use the following notations.

\paragraph*{Geometric notations}\

\begin{multicols}{2}
  \begin{description}[leftmargin=!, labelindent=0pt,labelwidth=0.5cm]
  \item[$\sphere^2$] unit sphere in $\IR^3$ 
  \item[$SO(3)$] group of rotations in $\IR^3$
  \item [$g_{A}, g_{B}$] orientation of grain $A$, $B$
  \item [$\Delta g = \mathtt{inv}(g_{A}) g_{B}$] misorientation
  \item [$\mathcal S_{A}, \mathcal S_{B}$] symmetry groups
  \item [$S_{A}, S_{B}$] symmetry elements in $\mathcal S_{A}$, $\mathcal S_{B}$  
  \item [$\vec n$] boundary normal in specimen coordinates
  \item [$\vec n_{A}  = \mathtt{inv}(g_{A}) \vec n$] boundary normal in crystal coordinates of grain $A$
  \item [$S_V$] surface area per unit volume
  \end{description}
\end{multicols}

\paragraph*{Statistical distributions}
\label{sec:density-functions}


\begin{description}[labelindent=0pt, widest={$\mathtt{BCD}_A(\Delta g, \vec n_{A})$},leftmargin=!, labelindent=0pt,labelwidth=2.3cm]
\item[$\mathtt{ODF}(g)$] orientation distribution function
\item[$\mathtt{ODF}_{A}(g_{A})$] orientation distribution function of $A$ grains
\item[$\mathtt{MDF}(\Delta g)$] misorientation distribution function
\item[$S_{V}(g_{A},\vec n,g_{B})$] surface area of boundary elements
  $(g_{A},\vec n,g_{B})$ per unit volume
\item[$\mathtt{BND}(\vec n)$] boundary normal distribution with
  respect to the specimen reference frame (specimen GBND)
\item[$\mathtt{BND}_{A}(\vec n_{A})$] boundary normal distribution with respect to the crystal reference frame of grain $A$ (crystal GBND)
\item[$\mathtt{BND}_{AB}(\vec n_{A})$]
same as above but with grain exchange symmetry
\item[$\mathtt{BCD}_A(\Delta g, \vec n_{A})$] boundary
  character distribution (GBCD)
\item[$\mathtt{BCD}_{AB}(\Delta g, \vec n_{A})$] same as above but with grain
  exchange symmetry
\item[$\mathtt{BCD}_{A}(\vec n_{A} | \Delta g)$] conditional GBCD 
\end{description}

\section{Statistical Descriptions of Boundary Networks}

\subsection{Boundaries}

From a purely geometrical perspective, grain boundary segments are most
commonly described by five parameters: three describing the misorientation
$\Delta g = \mathtt{inv(g_{A})g_{B}}$ between the adjacent crystal
orientations, and two describing the boundary normal $\vec n_{A}$ in crystal
coordinates, cf.~\cite{Sutton1995,Randle2010}. Those five parameters can be
accessed by means of 3D-EBSD \cite{Rohrer2010}, combinations of U-stage and
EBSD \cite{SaylorMorawiecRohrer2003_MgO5DOF}, diffraction
contrast tomography \cite{Ludwig:hx5063} or, statistically, by 2D EBSD data and stereological
approaches \cite{Saylor2004PlanarSections}. 

\begin{figure}
    \centering
\tdplotsetmaincoords{70}{120} 
\begin{tikzpicture}[tdplot_main_coords, scale=1]

  \coordinate (A) at (0,0,0);
  \coordinate (B) at (2.7,0,0);
  \coordinate (F) at (2,3,0); 
  \coordinate (D) at (0,3,0);
  \coordinate (A') at (0,0,3);
  \coordinate (B') at (3,0,3);
  \coordinate (E) at (2.5,3,3);
  \coordinate (D') at (0,3,3);
  
  \draw[fill=green!30, opacity=0.7] (A) -- (B) -- (B') -- (A') -- cycle;
  \draw[fill=green!30, opacity=0.7] (A) -- (D) -- (D') -- (A') -- cycle;
  \draw[fill=green!30, opacity=0.7] (A) -- (B) -- (F) -- (D) -- cycle;
  \draw[fill=green!30, opacity=0.7] (A') -- (B') -- (E) -- (D') -- cycle;
  \draw[fill=green!30, opacity=0.7] (B) -- (B') -- (E) -- (F) -- cycle;
  
  \coordinate (C)  at (5,0,0);
  \coordinate (C') at (5,0,3);
  \coordinate (H)  at (5,3,0);
  \coordinate (H') at (5,3,3);
  
  \draw[fill=orange!30, opacity=0.7] (B) -- (C) -- (C') -- (B') -- cycle;
  \draw[fill=orange!30, opacity=0.7] (B) -- (C) -- (H) -- (F) -- cycle;
  \draw[fill=orange!30, opacity=0.7] (B') -- (C') -- (H') -- (E) -- cycle;
  \draw[fill=orange!30, opacity=0.7] (C) -- (H) -- (H') -- (C') -- cycle;
  \draw[fill=orange!30, opacity=0.7] (F) -- (H) -- (H') -- (E) -- cycle;
  
  \draw[darkblue] (B) -- (B') -- (E) -- (F) -- cycle;
  
  \coordinate (M) at (2.75,1.5,1.5);
  \draw[->, very thick, darkblue] (M) -- ++(0.97, 0.22, -0.1) node[pos=0.8, below left]
  {$\vec n$};
  
  \node[darkgreen] at (1.375,1.5,3) {$g_A$};
  \node[darkred] at (4.375,1.5,3) {$g_B$};


  
\end{tikzpicture}
    
    \caption{Schematic presentation of a crystal boundary between two crystal of arbitrary crystal structures plotted within the sample reference frame}
    \label{fig:placeholder}
\end{figure}
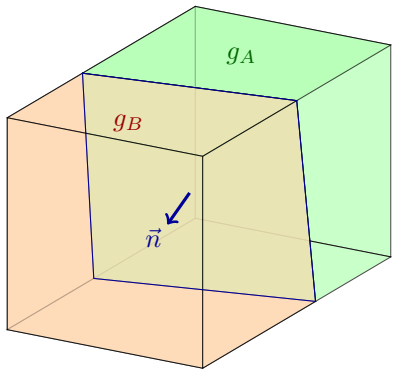

In this paper however, we employ the 8-parameter description pioneered by
Adams \cite{AdamsField1992}, who showed that five macroscopic DOF are
insufficient to characterize the boundary network fully. His formulation
extends the description to include the absolute orientations ($g_A$, $g_B$) of
both grains and the boundary normal $\vec n$ in specimen coordinates.
Ultimately, we describe a boundary element by the triple $(g_A,\vec n,g_B)$,
where the boundary normal $\vec n \in \sphere^2$ is chosen to point from grain
$A$ towards grain $B$, Fig.~\ref{fig:placeholder}. Here, the two orientations
are each described by 3-DOF and the boundary normal by 2-DOF. This 8-parameter
model should not be confused with microscopic extensions that combine
translational variables into an 8-DOF framework \cite{WINTER2025120968}.

Utilizing the grain orientations $g_A$ and $g_B$ the boundary normal $\vec n$ may also be represented with respect to the crystal reference frames of crystal $A$ and $B$ by $\vec n_A = \mathtt{inv}(g_A) \vec n$ and $\vec n_B = \mathtt{inv}(g_B) \vec n$ which then define the lattice planes $(hk\ell)_A$ in grain $A$ and $(hk\ell)_B$ in grain $B$ being in contact with each other.

For the statistical analysis of boundary elements we often rely on integrals
over the two-dimensional sphere $\sphere^2$, as well as over the rotation
group $SO(3)$. In the following we assume all those integrals to be normalized to one, i.e., 
\begin{equation*}
  \int_{SO(3)} 1 \d{g} = 1 \text{ and } \int_{\sphere^2} 1 \d{\vec n} = 1.
\end{equation*}
In other words, we treat all possible grain orientations, boundary directions, and misorientations as evenly weighted and scale such that the total adds up to one.

\subsection{Boundary Distribution Functions}

The simplest statistical description of a boundary network is its surface area
per unit volume $S_V$, c.f.~\cite{Underwood1970,VanderVoort2011}. Assuming
isotropic grain shape distributions this fraction is directly related to the
boundary length per unit area $L_{A} =  \frac{\pi}{4} S_{V}$ in
two-dimensional sections. Since the boundary surface between a grain $A$ and a grain $B$ is locally characterized by its normal vector $\vec n$ and the adjacent crystal orientations $g_A$ and $g_B$ it is natural to extend this number to a distribution $S_V(g_A,\vec n,g_B)$ that describes the surface area of boundaries as a function of $(g_A,\vec n,g_B)$, normalized such that the integral
\begin{equation*}
    \int_{SO(3)} \int_{\sphere^2} \int_{SO(3)} S_V(g_A,\vec n, \vec g_B) \d{g_A} \d{\vec n} \d{g_B}
    = S_V
\end{equation*}
over all possible boundary configurations coincides with the total surface by
volume fraction $S_V$. In other words the function $S_V(g_A,\vec n,g_B)$ does not only describe how much boundary area exists, but how this area is distributed among different boundary normals $\vec n$ and grain orientations $g_A$, $g_B$. The function $S_V(g_A,\vec n, g_B)$ was first introduced by Adams, cf.~\cite{AdamsField1992}.  It is important to understand that $S_V(g_A,\vec n,g_B)$ is a density function that is not supposed to be interpreted point-wise, but rather to be integrated over a certain subregion of boundary configurations. For example, the surface area per unit volume of boundaries with normals $\vec n$ distributed in a cone of angular tolerance $\delta$ around a reference  normal $\vec n_{\mathtt{ref}}$, i.e., $\angle(\vec n,\vec n_{\mathtt{ref}})<\delta$ and arbitrary grain orientation is the integral 
\begin{equation*}
  \int_{SO(3)} \int_{B_\delta(\vec n_\mathtt{ref})} \int_{SO(3)}
    S_V(g_A,\vec n,g_B) \d{g_A} d{\vec n} \d{g_B}
\end{equation*}
where $B_\delta(\vec n_\mathtt{ref}) \subset \sphere^2$ denotes the set of all directions $\vec n$ satisfying $\angle(\vec n,\vec n_{\mathtt{ref}})<\delta$.

All common boundary distributions, e.g. the 5-parameter grain boundary
character distribution (GBCD) and the 2-parameter boundary normal distribution
(GBND) can be derived from this 8-parameter distribution by specific choices
of subregions of integration. Starting with the 5-parameter grain boundary
character distribution $\mathtt{BCD}_A(\Delta g, \vec n_A)$, that is defined
as the relative surface area of boundaries with boundary normal $n_A$ in crystal coordinates and misorientation $\Delta g = \mathtt{inv}(g_A) g_B$. It is given by the integral
\begin{equation}
  \label{eq:gbcd}
  \mathtt{BCD}_A(\Delta g, \vec n_A)
 = \frac{1}{S_V} \int_{SO(3)}
    S_V(g_A,g_A \vec n_A, g_A \Delta g) \d{g_A}.
\end{equation}
Note that by dividing by $S_V$ the distribution $\mathtt{BCD}_A$ is normalized to multiples of the random distribution, i.e., the $\mathtt{BCD}_A$ integrates to $1$ and not to $S_V$.
As the GBCD is obtained by integrating over all grain orientations the full 8
DOF parameter space of $S_{V}$ reduces to 5 DOF for $\mathtt{BCD}_A$.

If we additionally neglect the information about the misorientation $\Delta g$
and are only interested in the distribution of the boundary normals $\vec n$
with respect to specimen reference frame, we arrive at the grain boundary
normal distribution $\mathtt{BND}(\vec n)$, which is given by the double integral
\begin{equation}
 \label{eq:gBND}
 \mathtt{BND}(\vec n)
 = \frac{1}{S_V} \int_{SO(3)} \int_{SO(3)}
    S_V(g_A,\vec n,g_B) \d{g_A} \d{g_B}.
\end{equation}
We will call this function \emph{specimen GBND} as it describes the
distribution of boundary normals with respect to the specimen reference frame
regardless of the misorientation. Similarly, we call the distribution function
\begin{equation}
 \label{eq:gBNDA}
 \begin{split}
   \mathtt{BND}_A(\vec n_A)
   &= \frac{1}{S_V} \int_{SO(3)} \int_{SO(3)}
   S_V(g_A, g_A \vec n_A, g_B) \d{g_A} \d{g_B}\\
   &= \int_{SO(3)} 
   \mathtt{BCD}_A(\Delta g,\vec n_A) \d{\Delta g}
 \end{split}
\end{equation}
that describes the surface area of boundaries with normal vector $\vec n_A$ with respect to the crystal reference frame  \emph{crystal GBND}. Note that the specimen GBND is in general an antipodal function, i.e., $\mathtt{BND}(\vec n) = \mathtt{BND}(-\vec n)$, whereas the crystal GBND does not necessarily possess this property.

From a mathematical perspective, all the above distributions are marginal
distributions of the 8-parameter distribution function $S_V(g_A,\vec
n,g_B)$. In other words, these distributions can be seen as reductions of the
full eight-parameter function, obtained by averaging over a subset of its
variables.  While it is mathematically possible to use the GBCD to derive the
\textit{crystal} GBND, it is impossible at the other hand to derive the
\textit{specimen} GBND without additional assumptions. The reason for this is
that the GBCD completely lacks any information about the alignment with
respect to the specimen reference frame.

\paragraph{Normalization}
All distributions $\mathtt{BND}$, $\mathtt{BND}_{A}$, $\mathtt{BCD}_{A}$
describe relative surface areas, i.e. are normalized such that they integrate
to $1$ and, hence, can be interpreted as multiples of the random
distribution. The surface per unit volume fraction for certain configurations
of boundary parameters is easily obtained by multiplying integrals of those
distribution with $S_V$.

It is important to distinguish between the GBCD distribution $\mathtt{BCD}_A(\Delta g,\vec n_A)$ and the \textit{conditional} GBCD
\begin{equation*}
  \mathtt{BCD}_A(\vec n_A | \Delta g)
    = \frac{\mathtt{BCD}_A(\Delta g, \vec n_A)}
    {\int_{\sphere^2} \mathtt{BCD}_A(\Delta g, \vec m_A) \d{\vec m_A}}
\end{equation*}
that gives the relative surface area of boundaries with normal $\vec n_A$
considering only boundaries with misorientation $\Delta g$. Both are
essentially the same distributions but with different normalization.
The GBCD is normalized such that the integral
\begin{equation*}
  \int_{\sphere^{2}} \mathtt{BCD}_A(\Delta g , \vec n_A) \d{\vec
    n_{A}}
  = \mathtt{MDF}(\Delta g) 
\end{equation*}
gives the relative surface area of
$\Delta g$ boundaries, allowing to quantitatively compare the GBCD for
different misorientations $\Delta g$. The conditional GBCD at the other hand is normalized to
\begin{equation*}
  \int_{\sphere^2} \mathtt{BCD}_A(\vec n_A | \Delta g) \d{\vec n_A} = 1,
\end{equation*}
which is better suited for analyzing the distribution of boundary normals with respect to a single misorientation.


\subsection{Symmetry properties of the distribution functions}

Several mathematically correct descriptions can represent the same physical
boundary. There are two situations: (i) crystal symmetry: symmetry-equivalent
and physically identical crystal orientations define the same type of
boundary 
and (ii) grain exchange symmetry, where we cannot distinguish between grain
$A$ and grain $B$.

Following we will investigate the consequences of those ambiguities for the boundary distributions in detail. (i) Since a crystal orientation $g_A$ is physically indistinguishable from symmetrically equivalent crystal orientations $g_A S_A$ for all symmetry elements $S_A \in \mathcal S_A$ of the corresponding symmetry group, we have the following equalities
\begin{align*}
   S_V(g_A, \vec n, g_B) 
   &= S_V(g_A S_A, \vec n, g_B S_B) \\
   \mathtt{BCD}_A(\Delta g, \vec n_A)
   &= \mathtt{BCD}_A(\mathtt{inv}(S_A) \Delta g S_B, S_A \vec n_A) \\
   \mathtt{BND}_A(\vec n_A)
   &= \mathtt{BND}_A(S_A \vec n_A)
\end{align*}
for all symmetry elements $S_A \in \mathcal S_A$ and $S_B \in \mathcal S_B$. Note that the crystal GBND $\mathtt{BND}_A(\vec n_A)$  possesses crystal symmetry with respect to $\vec n_A$, but the crystal GBCD $\mathtt{BCD}_A(\Delta g, \vec n_A)$ does not.

Secondly, if grain $A$ and grain $B$ cannot be distinguished, i.e. both
belong to the same phase and we can only arbitrarily assign the label $A$ to
the grain on the one side of the boundary and label $B$ to the grain on the
other side of the boundary, swapping grains should not change the
distributions. This implies the equalities
\begin{align*}
   S_V(g_A, \vec n, g_B) 
   &= S_V(g_B, -\vec n, g_A) \\
   \mathtt{BCD}_{AB}(\Delta g, \vec n_A)
   &= \mathtt{BCD}_{AB}(\mathtt{inv}(\Delta g), -\mathtt{inv}(\Delta g) \vec n_A) \\
   \mathtt{BND}_{AB}(\vec n_A)
   &= \mathtt{BND}_{AB}(-\vec n_A).
\end{align*}
Here we used the notations $\mathtt{BND}_{AB}$ and $\mathtt{BCD}_{AB}$ to
stress the ambiguity between $A$ and $B$ grains. 

Cases which do not involve grain exchange symmetry include boundaries between unlike phases or in cases where a specific order can be assigned. The latter may for example cover twin boundaries where the twin and the host can be clearly distinguished.

\subsection{Determining Boundary Distribution Functions from 3D-Microstructures}

Determining grain boundary distributions from a measured microstructure is a
challenging problem. If only two-dimensional sections are available,
stereological methods must be applied,
cf. \cite{LarsenAdams2004NewStereology,Saylor2004PlanarSections}. For fully
three-dimensional measurements, several experimental and reconstruction
challenges must be addressed, including section alignment, sample drift,
anisotropic voxels, and smoothing of voxelized boundary surfaces; see, e.g.,
\cite{Rohrer2011,Rollett2007}. In this section we ignore all these challenges and assume a perfect three-dimensional description of the microstructure. More precisely, we assume that the grain boundaries are given as faces with normal directions $\vec n^k$, $k=1,\ldots,K$, and surface areas $s_k$, pointing from grains with orientation $g_A^k$ toward grains with orientation $g_B^k$. Denoting the total surface area by
\begin{equation*}
  S_{\text{total}} = \sum_{k=1}^K s_k
\end{equation*}
and the total volume by $V$ we obviously have $S_V = S_{\text{total}}/V$.

\begin{figure}
  \begin{subfigure}{0.26\linewidth}
    \includegraphics[width=\linewidth]{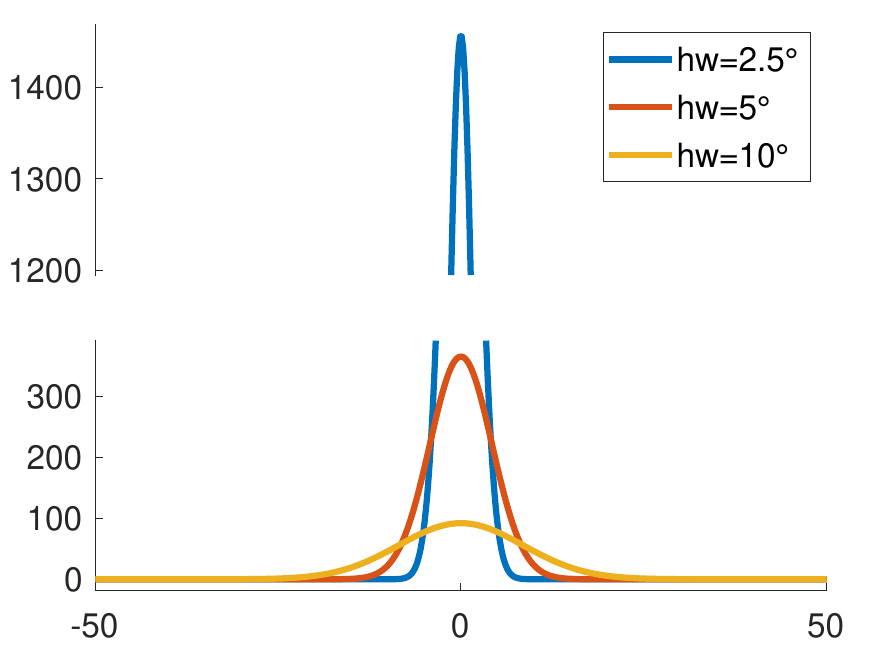}
    \caption{the kernel function $\psi$}\label{fig:psi}
  \end{subfigure}
  \hfill
  \begin{subfigure}{0.20\linewidth}
    \includegraphics[width=\linewidth]{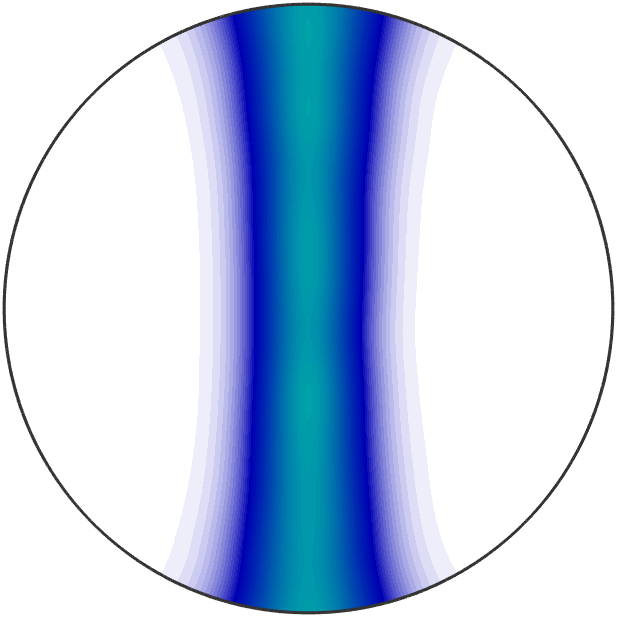}
    \caption{$\mathtt{BND}^{*}$, hw$=10^\circ$}\label{fig:gbnd10}
  \end{subfigure}
    \hfill
  \begin{subfigure}{0.20\linewidth}
    \includegraphics[width=\linewidth]{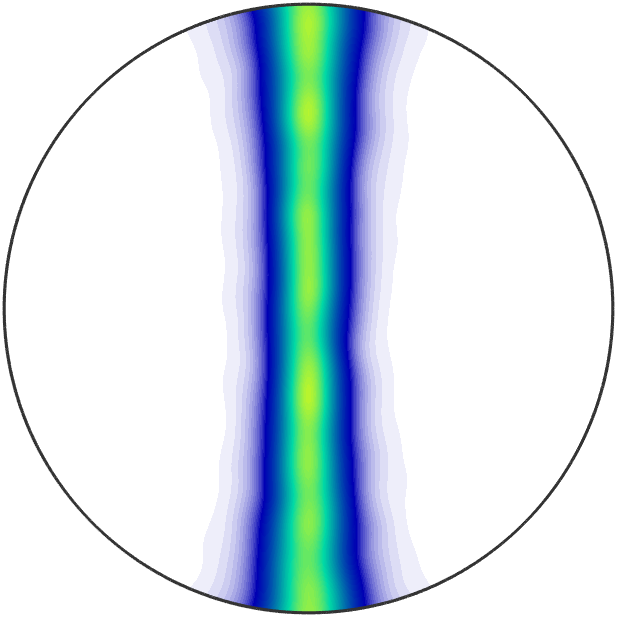}
    \caption{$\mathtt{BND}^{*}$, hw$=5^\circ$}\label{fig:gbnd5}
  \end{subfigure}
    \hfill
  \begin{subfigure}{0.22\linewidth}
    \includegraphics[width=\linewidth]{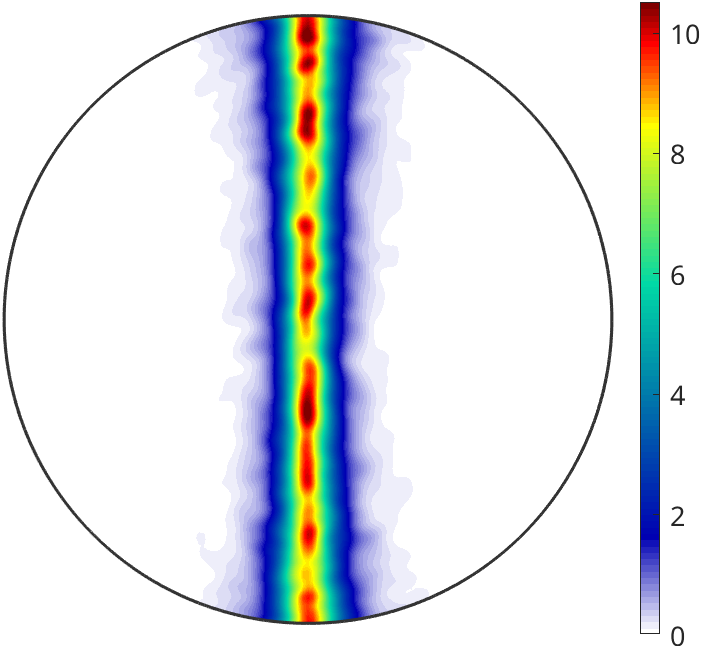}
    \caption{$\mathtt{BND}^{*}$, hw$=2.5^\circ$}\label{fig:gbnd25}
  \end{subfigure}    

  \caption{Spherical kernel functions $\psi$ with different halfwidth
    \subref{fig:psi}. The specimen GBNDs of the microstructure in
    \ref{fig:grainMacroUniform} were computed using these kernel functions
    \subref{fig:gbnd10}-\subref{fig:gbnd25}.  }\label{fig:gbnd}
  
\end{figure}

\paragraph{Estimating the specimen GBND}
Our method of estimating the boundary distribution functions is kernel density
estimation \cite{Silverman1986,HallWatsonCabrera1987, Hielscher2013}. To this
end we fix a radial kernel function on the sphere $\psi(\vec n) = \Psi(\vec n
\cdot \vec z)$ as depicted in a cross section in Fig.~\ref{fig:psi} that is
normalized to $\int_{\mathbb S^{2}} \psi(\vec n)=1$.
Taking the surface weighted average of those kernel functions centered at face normals $\vec n^k$ we define the kernel density estimator of the specimen GBND as
\begin{equation}
  \mathtt{BND}^{*}(\vec n)
  = \frac{1}{2S_{\text{total}}} 
  \sum_{k=1}^K  s_k \left(\Psi(\vec n \cdot \vec n^k)
  +\Psi(-\vec n \cdot \vec n^k))    
    \right).
\end{equation}
Note that averaging over $\pm \vec n \cdot \vec n^k$ ensures that
the estimated GBND is antipodal. As for all kernel density estimators, the estimated specimen GBND heavily depends on the chosen kernel halfwidth. In Fig.~\ref{fig:gbnd} the influence of the kernel halfwidth is illustrated when estimating the GBND of the simulated microstructure from Fig.~\ref{fig:grainMacroUniform}. Clearly a too sharp kernel function leads to oscillatory behavior, cf.~Fig.~\ref{fig:gbnd25}, while a too wide kernel function results in oversmoothing, cf.~Fig.~\ref{fig:gbnd10}.

\begin{figure}
  \begin{subfigure}{0.3\linewidth}
    \includegraphics[width=\linewidth]{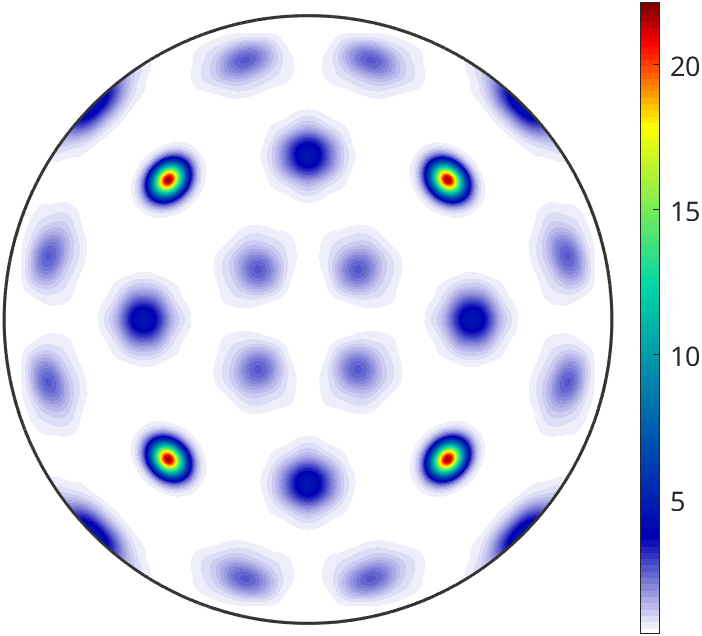}
    \caption{$\mathtt{BND}_{AB}^{*}$}\label{fig:gbndA}
  \end{subfigure}
  \hfill
  \begin{subfigure}{0.3\linewidth}
    \includegraphics[width=\linewidth]{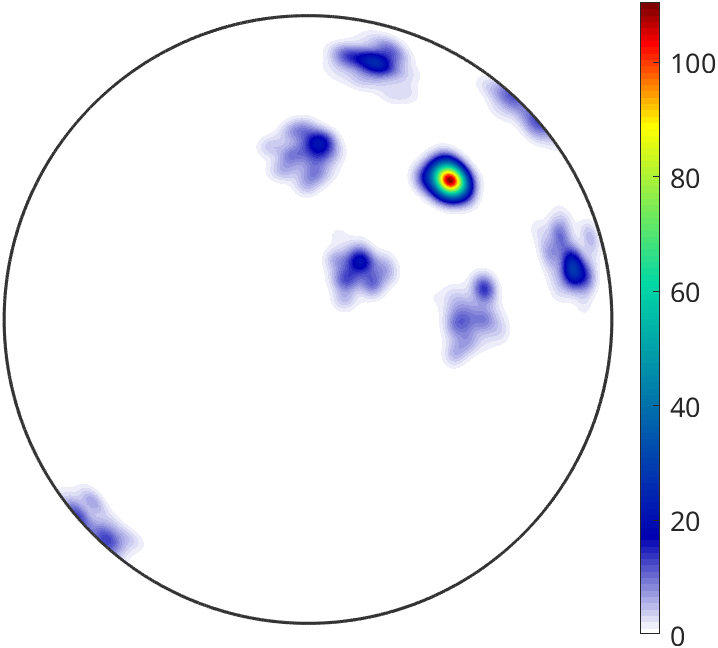}
    \caption{$\mathtt{BCD}_{AB}^{*}$}\label{fig:gbcdnoSym}
  \end{subfigure}
    \hfill
  \begin{subfigure}{0.3\linewidth}
    \includegraphics[width=\linewidth]{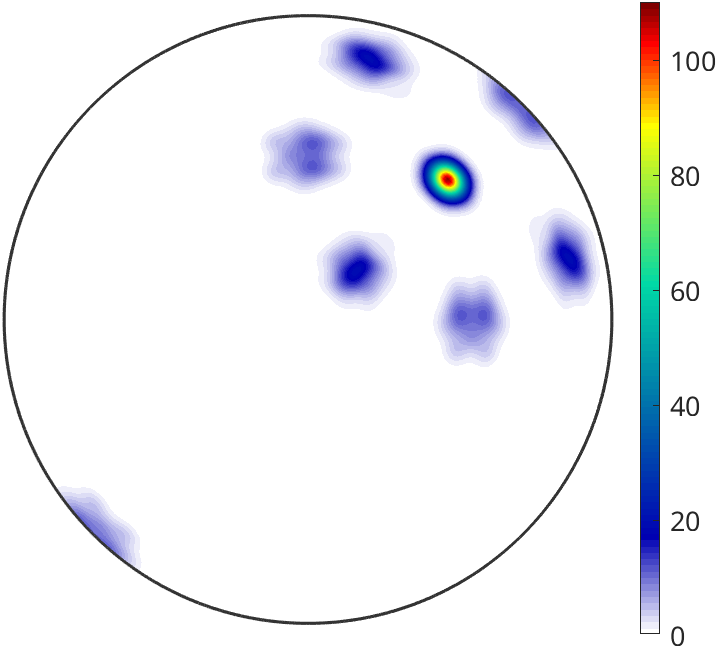}
    \caption{symmetrized $\mathtt{BCD}_{AB}^{*}$}\label{fig:gbcdSym}
  \end{subfigure}
    
  \caption{Crystal GBND \subref{fig:gbndA} and GBCD estimated from the
    simulated microstructure in Fig.~\ref{fig:grainDynamicUniform}. For this
    visualization not the entire data set has been used but only $10\,000$
    randomly sampled surface elements out of $320\,410$. This reduction reveals
    the difference between the non-symmetrized GBCD \subref{fig:gbcdnoSym} and
    the symmetrized GBCD visible.}\label{fig:gbcd}
  
\end{figure}

\paragraph{Estimating the crystal GBND}
For the crystal GBND we make use of the accompanying orientations $g_A^k$ to
transform the  boundary normals $\vec n_{k}$ into the crystal reference frame,
i.e., $\vec n_{A}^{k}=\mathtt{inv}(g_{A}^{k}) \vec n_{k}$. Symmetrization with
respect to $\mathcal S_A$ gives 
\begin{equation*}
  \mathtt{BND}_A^{*}(\vec n_A)
  = \frac{1}{\abs{\mathcal S_A}} \sum_{S_A \in \mathcal S_A}
  \frac{1}{S_{\text{total}}}\sum_{k=1}^K  s_k 
    \Psi(\vec n_A \cdot S_A \vec n_{A}^k),
\end{equation*}
where $\abs{\mathcal S_{A}}$ denotes the number of symmetry elements in
$\mathcal S_A$. The crystal GBND computed from the simulated $\Sigma3$
twinning microstructure, cf. Fig.~\ref{fig:grainDynamicUniform}, is depicted
in Fig.~\ref{fig:gbndA}.

\paragraph{Estimating the GBCD}
Further work is required to obtain a reliable estimate of the GBCD. Given a misorientation $\Delta g$ we first determine for each pair of neighboring orientations $g_A^k$, $g_B^k$ symmetry elements $S_A^k \in \mathcal S_A$ and $S_B^k \in \mathcal S_B$ such that the misorientation angle 
\begin{equation}
  \label{eq:minProblem}
  \omega_k = \angle(\Delta g, \mathtt{inv}(g_A^k S_A^k) g_B^k S_B^k)
\end{equation}
with respect to $\Delta g$ is minimal. Next we use those specific symmetry
elements in conjunction with the grain orientations to transform the surface
normals $\vec n^{k}$ into crystal reference frame, i.e., $\vec n_{A}^{k} =
\mathtt{inv}(g_A^k S_A^k) \vec n^{k}$. Eventually, we obtain the estimators
\begin{align*}
  \mathtt{BCD}_A^{*}(\Delta g, \vec n_A)
  &= \frac{1}{S_{\text{total}}} \sum_{k=1}^K  s_k 
    \Phi(\cos \omega_k/2) \,
    \Psi(\vec n_A \cdot \vec n_{A}^k)
    \bigg/ \int_{0}^{\pi} \Phi(\cos \omega/2) \sin^{2}(\omega/2) \d{\omega} ,\\
  \mathtt{BCD}_A^{*}(\vec n_A | \Delta g)
  &= \sum_{k=1}^K  s_k 
    \Phi(\cos \omega_k/2) \,
    \Psi(\vec n_A \cdot \vec n_{A}^k)
    \bigg/
    \sum_{k=1}^K s_k 
    \Phi(\cos \omega_k/2),
\end{align*}
where the kernel function $\Phi$ models the smoothing with respect to the
misorientation angle. Again, choosing a reasonable half-width is crucial.  In
this paper we will always use the same kernel for both misorientation
and boundary normals, i.e. $\Psi = \Phi$.


\paragraph{About the uniqueness of the symmetry elements $S_A^k$}
It should be noted that for specific choices of misorientations $\Delta g$ the symmetry elements $S_A^k$, $S_B^k$ minimizing Eq.~\eqref{eq:minProblem} are not unique. This happens exactly for those misorientations, where there are symmetry elements $S_A$, $S_B$ satisfying 
\begin{equation}
  \label{eq:intertwin}
  \Delta g S_B = S_A \Delta g.
\end{equation}
A typical example of such a misorientation is $\Sigma 3\left<111\right>$ twinning in fcc or bcc  materials. In this case the symmetry elements of 432 that satisfy Eq.~\eqref{eq:intertwin} are precisely the elements of the $\left<111\right>$
 threefold axis, i.e., rotations by $\pm 120^{\circ}$ about $\left<111\right>$.

Generally, the set of all symmetry elements $S_A$, $S_B$ satisfying Eq.~\eqref{eq:intertwin} form subgroups of $\mathcal S_A$ and $\mathcal S_B$ which we denote by $\mathcal S_A(\Delta g)$ and $\mathcal S_B(\Delta g)$. To make our density estimator aware of the additional symmetry $\mathcal S_A(\Delta g)$ we simply average over all those symmetry elements 
\begin{equation*}
\label{eq:bcdEstSym}
  \mathtt{BCD}^*(\Delta g| \vec n_A)
  = \frac{1}{\abs{\mathcal S_A(\Delta g)}}
  \sum_{S_A \in \mathcal S_A(\Delta g)} \sum_{k=1}^K  s_k 
  \Phi(\cos \omega_k/2) \,
  \Psi(\vec n_A \cdot S_A \vec n_{A}^k)
  \bigg/
    \sum_{k=1}^K s_k 
    \Phi(\cos \omega_k/2).
\end{equation*}
In the remainder of this paper we will always use this symmetrized version. The
difference between the symmetrized and not symmetrized GBCD is illustrated in
Fig.~\ref{fig:gbcd} for the twinning microstructure depicted in Fig.~\ref{fig:grainDynamicUniform}.

\paragraph{Numerical Implementation}
All estimators introduced above involve large sums of radial functions on the
sphere and on the orientation space. A direct evaluation of these sums becomes
computationally infeasible in practical applications, where millions of grain
boundary segments may be involved. Efficient evaluation is enabled by fast
Fourier-based algorithms on the sphere and on $SO(3)$, see, e.g.,
\cite{Kunis2003,Hielscher2010,Hielscher2013,Hielscher2026}. A brief outline of the underlying ideas is
given in Appendix~\ref{sec:fast-summ-spher}. These algorithms form the
computational foundation of the implementations of the above estimators in
MTEX \cite{MTEX}.

\section{Model Distributions for Specific Boundary Formation Processes}

Although the 8-parameter boundary distribution function $S_V(g_A,\vec n,g_B)$ gives the most complete statistical description of the boundary network it is very hard to visualize, to analyze, or to determine experimentally due to its high dimensional parameter space. For that reason, we introduce for specific scenarios simplified model distributions where some of the parameters are statistically independent. 


We shall distinguish two scenarios. In the first scenario, covered in Section~\ref{sec:macr-init-bound}, the formation of preferred boundary normal directions is driven by a macroscopic, sample scale process independent of the crystal orientations, e.g. deformation or grain growth.
In the second scenario, covered in Section \ref{sec:crystDriven}, the formation of preferred boundary normal directions is driven by some crystallographic process, such as normal or abnormal grain growth or twinning, and unrelated to their orientation with respect to the sample.

The two model classes should be understood as idealized end-members. Real microstructures may combine both macroscopic alignment effects and crystallographic selection effects.

\subsection{Macroscopically driven boundary networks}
\label{sec:macr-init-bound}

In this section we assume that the formation of preferred grain boundary normal directions is a consequence of a macroscopic process and independent of the local lattice orientations. 

Situations that may produce a macroscopically driven boundary distribution include, for example:
\begin{itemize}
\item A deformed material composed of flattened or stretched grains that attained their shape without the development of crystallographic texture, e.g. a material deforming by diffusion or dissolution--precipitation creep leading to anisotropic, aligned grain shapes without the development of texture \cite{LIFSHITZ1963, Rutter1976PressureSolution, FordWheelerMovchan2002}.
\item Fibrous vein growth which frequently lacks growth competition and hence also lacks the development of a crystallographic texture \cite{URAI1991, MEANS2001, Bons1997Exp} and generally a microstructural evolution governed by geometric confinement for example evident in confined crystal growth \cite{Kohler2022,Rehn2019,Spear2024}.
\item These observations are relevant for situations in which the grain boundary network is primarily controlled by macroscopic processes rather than local crystallography, for example by finite strain or deformation. One example is a von Mises-compliant material deforming by dislocation glide, where the grain shape—and thus the boundary network—is governed by the imposed strain, while a crystallographic texture develops depending on the active slip systems
\cite{Chin1973,Haase2013}. Similarly, in deforming and dynamically recrystallizing materials, grain shapes may deviate from the finite strain and instead reflect the instantaneous deformation \cite{REE1991,SCHMID1987}.
Comparable situations arise during combined grain boundary sliding and dislocation activity \cite{Ferreira2021}. In all these cases, the boundary network is macroscopically driven, even though a non-uniform ODF may be present.
\end{itemize}

\paragraph{Mathematical model of macroscopically driven boundary  networks}
The key statistical assumption for macroscopically driven boundaries  is that the distribution of boundary normals $\vec n$
in the specimen reference frame are statistically independent of the adjacent
grain orientations $g_A$ and $g_B$. This is modeled by factorizing the eight-parameter boundary distribution function
\begin{equation}
\label{eq:svMacro}
  S_V(g_A,\vec n,g_B)
  = S_V \cdot \mathtt{ODF}_{AB}(g_{A},g_{B}) \cdot 
 \mathtt{BND}(\vec n)
\end{equation}
into a product of a \textit{two-grain orientation distribution function}
$\mathtt{ODF}_{AB}(g_{A},g_{B})$ and the specimen GBND $\mathtt{BND}(\vec
n)$. Important examples for such a two-grain orientations distribution function are
$\mathtt{ODF}_{AB}(g_{A},g_{B}) = \mathtt{ODF}(g_{A}) \cdot \mathtt{MDF}(\mathtt{inv}(g_{A}) g_{B})$ or
$\mathtt{ODF}_{AB}(g_{A},g_{B}) = \mathtt{ODF}(g_{A}) \cdot \mathtt{ODF}(g_{B})$.

We now ask ourselves whether an anisotropic distribution of the boundary
normals with respect to the specimen reference frame, i.e. $\mathtt{BND} \ne
1$, may result in an anisotropic distribution of boundary normals with respect
to the crystal reference frame. To answer this question we use
Eq.~\eqref{eq:gBNDA} and obtain for the crystal GBND
\begin{equation}
  \begin{split}
    \label{eq:bnd2bndA}        
  \mathtt{BND}_{A}(\vec n_{A})
  &= \int_{SO(3)} \int_{SO(3)} \mathtt{ODF}_{AB}(g_{A},g_{B}) \cdot
    \mathtt{BND}(g_{A} \vec n_{A}) \d{g_{B}} \d{g_{A}} \\
  &= \int_{SO(3)} \mathtt{ODF}(g_{A}) \cdot
    \mathtt{BND}(g_{A} \vec n_{A}) \d{g_{A}}
  = \mathtt{ODF} \circledast \mathtt{BND}(\vec n_{A}).
  \end{split}
\end{equation}  
This equation establishes a quantitative relation between the specimen GBND, the ODF and the crystal GBND in macroscopically driven boundary networks. More specifically, the ODF and the specimen GBND determine
a corresponding crystal GBND, namely their spherical convolution, see Appendix~\ref{sec:spher-conv}. The spherical convolution can be interpreted as averaging the specimen GBND over all crystal orientations weighted by the ODF. 

This leads to the following consequences for macroscopically driven boundary
networks:
\begin{enumerate}
\item In an untextured material, i.e.\ $\mathtt{ODF} = 1$, the crystal GBND is
  uniform as well, i.e.\ $\mathtt{BND}_{A} = 1$. \label{item:1}
\item In a textured material, i.e.\ $\mathtt{ODF} \ne 1$, a non-uniform specimen GBND, i.e.\ $\mathtt{BND} \ne 1$, may result in a non-uniform crystal GBND, $\mathtt{BND}_{A} \ne 1$.
\item Since convolution acts as a smoothing operation, the crystal GBND is typically less sharp than the specimen GBND.
\end{enumerate}

This relationship between crystal and specimen GBND is crucial for interpreting a measured crystal GBND in the situations mentioned at the beginning of this sections. In the case of a material without a crystallographic texture any non-uniform crystal GBND indicates the presence of a boundary selection mechanism which is influenced by the crystallography of the material. For textured materials the observation of a non-uniform crystal GBND does not necessarily imply the presence of such a mechanism. Instead one has to compare the measured crystal GBND with the 
texture-induced crystal GBND $\mathtt{ODF} \circledast \mathtt{BND}$. If those distributions differ significantly one can again conclude on the presence of an additional crystallographic mechanism.

\begin{figure}
  \begin{subfigure}{0.34\linewidth}
    \includegraphics[width=\linewidth]{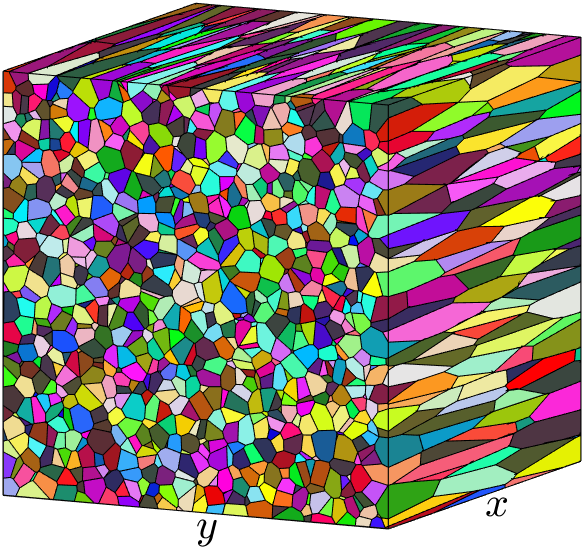}
    \caption{untextured microstructure}\label{fig:grainMacroUniform}
  \end{subfigure}
  \hfill
  \begin{subfigure}{0.26\linewidth}
    \includegraphics[width=\linewidth]{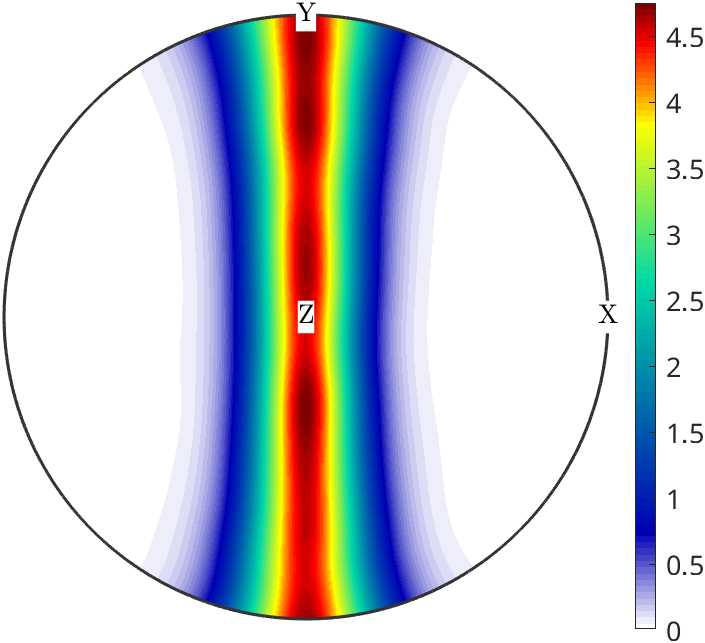}
    \caption{measured $\mathtt{BND}^{*}$}\label{fig:macroBND}
  \end{subfigure}
  \hfill
  \begin{subfigure}{0.34\linewidth}
    \includegraphics[width=\linewidth]{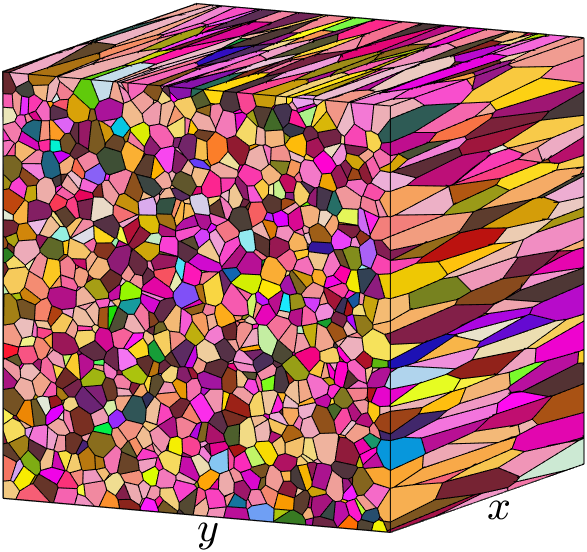}
    \caption{textured microstructure}\label{fig:grainMacroTex}
  \end{subfigure}

  \bigskip
  
  \begin{subfigure}{0.24\linewidth}
    \includegraphics[width=\linewidth]{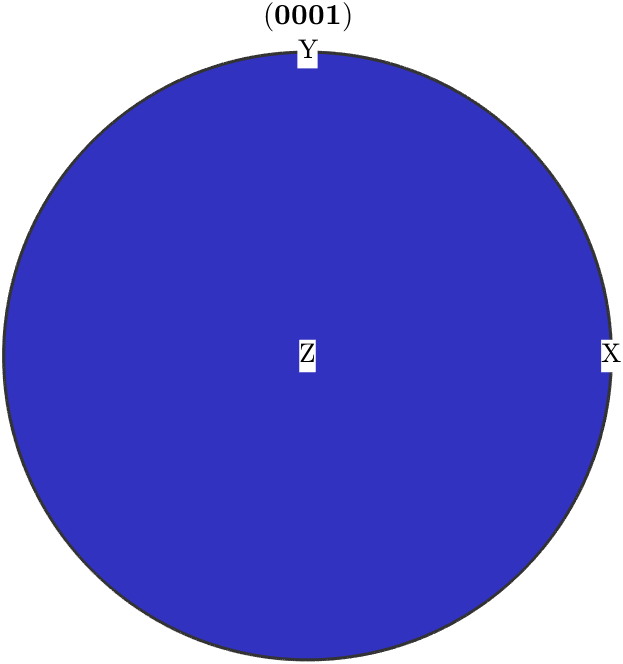}
    \caption{initial $\mathtt{ODF}$}\label{fig:pfMacro}
  \end{subfigure}
  \begin{subfigure}{0.26\linewidth}
    \includegraphics[width=\linewidth]{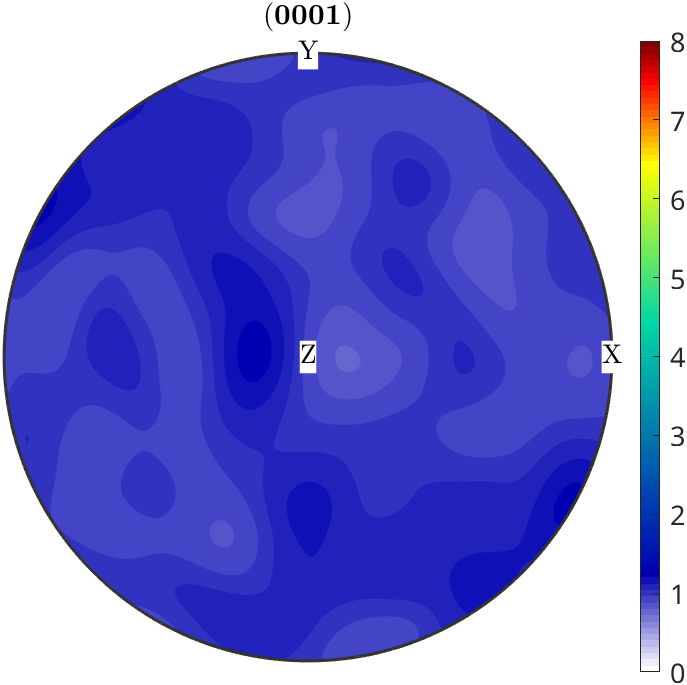}
    \caption{measured $\mathtt{ODF}^{*}$}\label{fig:pfMacroSim}
  \end{subfigure}\hfill
  \begin{subfigure}{0.24\linewidth}
    \includegraphics[width=\linewidth]{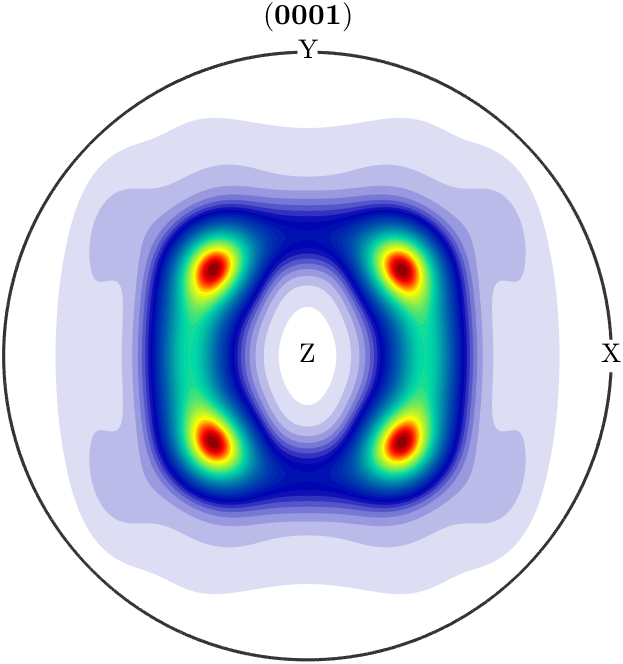}
    \caption{initial $\mathtt{ODF}$}\label{fig:pfMacroTex}
  \end{subfigure}
  \begin{subfigure}{0.24\linewidth}
    \includegraphics[width=\linewidth]{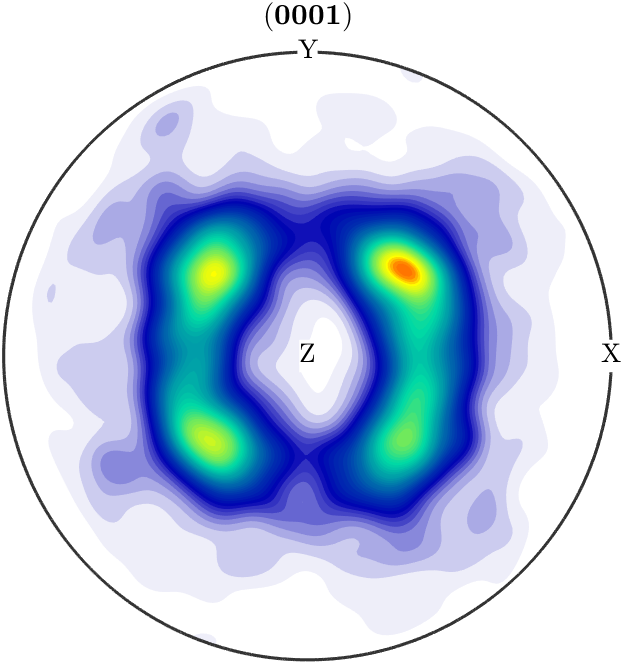}
    \caption{measured $\mathtt{ODF}^{*}$}\label{fig:pfMacroTexSim}
  \end{subfigure}

  \bigskip
  
  \begin{subfigure}{0.24\linewidth}
    \includegraphics[width=\linewidth]{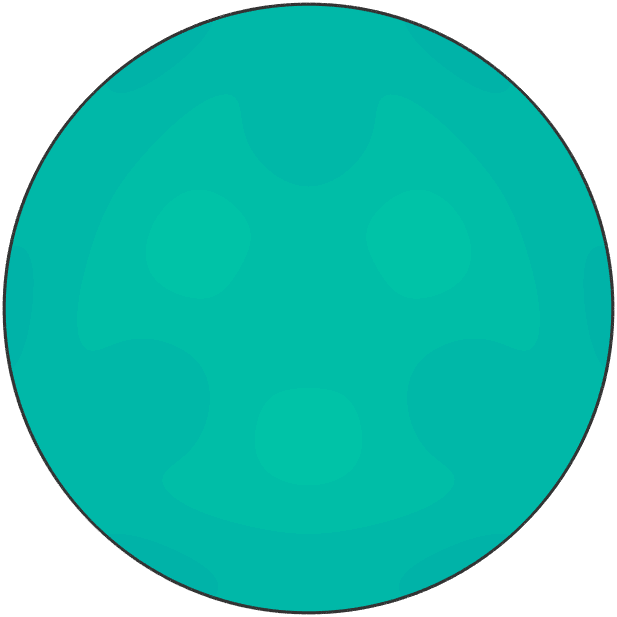}
    \caption{predicted $\mathtt{BND}_{A}$}\label{fig:bndAMacroUniform}
  \end{subfigure}
  \begin{subfigure}{0.26\linewidth}
    \includegraphics[width=\linewidth]{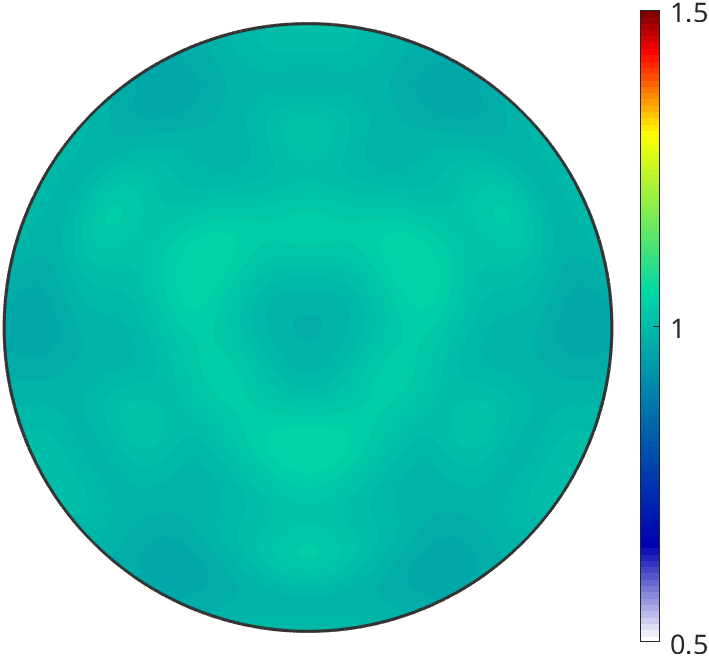}
    \caption{measured $\mathtt{BND}_{A}^{*}$}\label{fig:bndAMacroUniformSim}
  \end{subfigure}\hfill
  \begin{subfigure}{0.24\linewidth}
    \includegraphics[width=\linewidth]{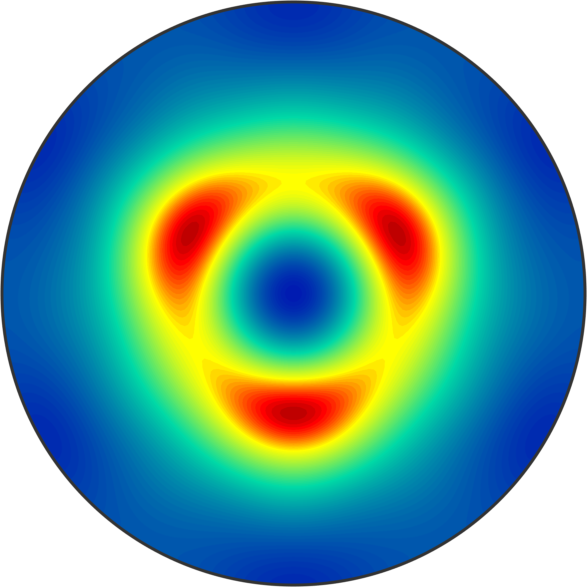}
    \caption{predicted $\mathtt{BND}_{A}$}\label{fig:bndAMacroTex}
  \end{subfigure}
  \begin{subfigure}{0.24\linewidth}
    \includegraphics[width=\linewidth]{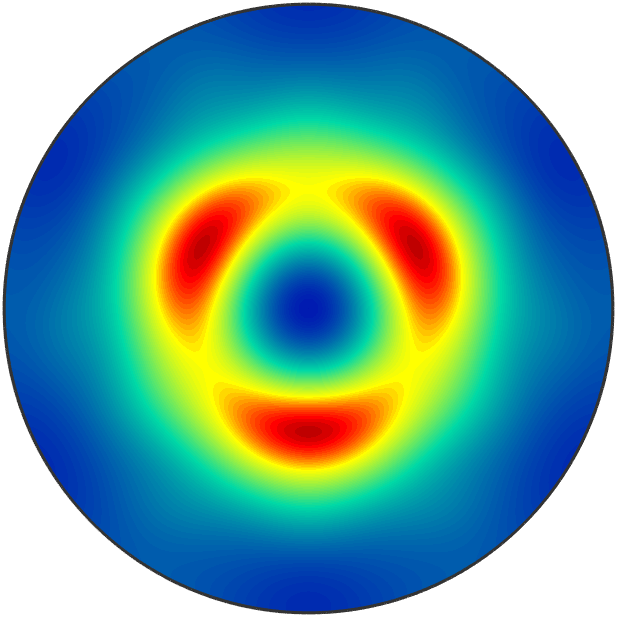}
    \caption{measured $\mathtt{BND}_{A}^{*}$}\label{fig:bndAMacroTexSim}
  \end{subfigure}
  
  \caption{Simulated Quartz microstructure with a preferred grain elongation
    in $x$-direction resulting in the specimen GBND \subref{fig:macroBND}. In
    \subref{fig:grainMacroUniform} the grain orientations have been sampled
    from the uniform distribution $\mathtt{ODF}=1$ \subref{fig:pfMacro} and in
    \subref{fig:grainMacroTex} from the sharp orthotropic ODF depicted in
    \subref{fig:pfMacroTex}. The actual textures of the simulated
    microstructures are depicted in \subref{fig:pfMacroSim} and
    \subref{fig:pfMacroTexSim}, respectively. The resulting crystal GBNDs are
    depicted in \subref{fig:bndAMacroUniformSim} and
    \subref{fig:bndAMacroTexSim} for the untextured and textured
    microstructure, respectively. The theoretical predictions
    $\mathtt{ODF}^{*} \circledast \mathtt{BND}^{*}$, computed by
    Eq.~\eqref{eq:bnd2bndA} using the measured ODFs \subref{fig:pfMacroSim},
    \subref{fig:pfMacroTexSim} and the measured GBND \subref{fig:macroBND},
    are depicted in \subref{fig:bndAMacroUniform} and
    \subref{fig:bndAMacroTex}.}\label{fig:Macro}
\end{figure}

\paragraph{Simulated example of a macroscopically driven microstructure}
Let us illustrate the above findings with a simulated three-dimensional
microstructure. Firstly, we use Neper to simulate a microstructure of 4000
Quartz grains with a preferred elongation in x-direction and uniform ODF as
depicted in Fig.~\ref{fig:grainMacroUniform}. In Fig.~\ref{fig:macroBND} we
observe a specimen GBND with a very pronounced fiber perpendicular to the
preferred grain elongation direction $\vec x$. In contrast, the observed
crystal GBND in Fig.~\ref{fig:pfMacroTexSim} is almost perfectly uniform as
predicted in point \ref{item:1}.

In a second experiment we populated the grains with orientations, randomly
drawn from a sharp ODF as depicted by its c-axis pole figure in
Fig.~\ref{fig:pfMacroTex}. The resulting microstructure together with the
estimated c-axis pole figure and the estimated and predicted crystal GBND are
displayed on the right hand side of Fig.~\ref{fig:Macro}. Since the geometric
structure of the grain boundaries remained exactly the same, both
microstructures share the same specimen GBND as depicted in
Fig.~\ref{fig:macroBND}. However, the measured crystal GBND, Fig.~\ref{fig:bndAMacroTexSim}, of the textured
microstructure shows clear maxima for the planes $[10\bar{1}2]$,
almost perfectly matching the predicted crystal
GBND $\mathtt{ODF} \circledast \mathtt{BND}$, Fig.~\ref{fig:bndAMacroTex},
computed solely from the ODF and the specimen GBND.

\subsection{Crystallographically driven boundary networks.}
\label{sec:crystDriven}

In contrast to the previous section we now assume that the boundary network is purely governed by crystallographic mechanisms. Those we assume to be described by the grain boundary character distribution $\mathtt{BCD}(\Delta g,\vec n_A)$, independently of the specimen reference frame. Multiple processes and mechanisms that can govern a grain-boundary network have been identified: energy minimization, kinetic control and kinematic control \cite{Rohrer2014,Marquardt2015}. 

Examples of such mechanisms include
\begin{itemize}
\item the formation of growth twins during crystallization or annealing
processes \cite{ChristianMahajan1995Twins},

\item polycrystalline aggregates formed by synneusis of euhedral crystals in
a melt. In this case crystals grow with a well-defined crystal habit and
preferentially attach along their dominant crystallographic planes
\cite{Wieser2019, Wieser2020},

\item topotactic growth interfaces occurring during solid-state reactions or
phase transformations where orientation relationships constrain the boundary
geometry \cite{Cahn1964Interfaces,PorterEasterling2009PhaseTransformations},

\item randomly oriented crystals without crystallographic texture but with
euhedral grain shapes controlled by anisotropic grain boundary energies,
leading to preferred crystallographic boundary planes
\cite{Herring1951SurfaceEnergy,Wulff1901CrystalShape}, that may change as a function of external conditions (temperature, pressure...), causing complexion transitions \cite{Austin2025}.

\end{itemize}

\paragraph{Mathematical model of crystallographically driven boundary
  networks}
Assuming statistical independence between the grain
orientations $g_A$ and the grain boundary characteristics
$(\Delta g, \vec n_A)$ we obtain the following model of the 8-parameter
boundary distribution function
\begin{equation}
    \label{eq:SvCrystal}
    S_V(g_A,\vec n,g_B)
    = S_V \cdot \mathtt{ODF}_{A}(g_{A}) \cdot
      \mathtt{BCD}_A(\mathtt{inv}(g_{A}) g_B, \mathtt{inv}(g_{A}) \vec n)      
\end{equation}
as a product of the ODF $\mathtt{ODF}_{A}$ and the GBCD $\mathtt{BCD}_A$. Here, grain exchange symmetry has not been assumed and $\mathtt{ODF}_{A}$ denotes the distribution of the $A$-grains. Using the misorientation distribution function
\begin{equation}
  \mathtt{MDF}(\Delta g)
  = \int_{\sphere^2} \mathtt{BCD}_A(\Delta g, \vec n_A) \d{\vec n_A}
\end{equation}
we obtain for the ODF of the $B$-grains
\begin{align*}
  \mathtt{ODF}_{B}(g_{B})
  &= \int_{SO(3)} \int_{\sphere^2} \mathtt{ODF}_{A}(g_{A}) \cdot 
  \mathtt{BCD}_A(\mathtt{inv}(g_{A}) g_B, \mathtt{inv}(g_{A}) \vec n) \d{\vec n} \d{\vec g_A}\\
  &= \int_{SO(3)} \mathtt{ODF}_{A}(g_{A}) \cdot 
  \mathtt{MDF}(\mathtt{inv}(g_{A}) g_B) \d{\vec g_A}
  = \mathtt{MDF} \ast \mathtt{ODF}_{A}(g_{B}),
\end{align*}
i.e., the ODF of the $B$ grains is given by the rotational convolution (see Appendix~\ref{sec:spher-conv})
of ODF of the $A$ grains with the misorientation distribution function
$\mathtt{MDF}$. Accordingly, the combined ODF of $A$ and $B$ grains resolves to
$\mathtt{ODF} = \frac{1}{2}(\mathtt{ODF}_{A} + \mathtt{MDF} \ast
\mathtt{ODF}_{A})$.

Similarly, our model allows for distinct crystal GBNDs with respect to the A-grains $\mathtt{BND}_A$ and with respect to the B-grains $\mathtt{BND}_{B}$. Using Eq.~\eqref{eq:gBNDA} we obtain
\begin{equation}
  \label{eq:bndACrystal}
  \begin{split}
  \mathtt{BND}_A(\vec n_A)
  &= \int_{SO(3)} \mathtt{BCD}_A(\Delta g, \vec n_A) \d{\Delta g}\\
  \mathtt{BND}_{B}(\vec n_B)
  &= \int_{SO(3)} \mathtt{BCD}_A(\Delta g, -\Delta g \, \vec n_B) \d{\Delta g}
  \end{split}  
\end{equation}
In case of grain exchange symmetry the resulting crystal GBND is simply the
average $\mathtt{BND}_{AB} = \tfrac12(\mathtt{BND}_A + \mathtt{BND}_{B})$.

The central result of this section is an analogous, quantitative relationship between  specimen GBND, the crystal GBND and the ODF as for macroscopically driven boundary networks in Eq.~\ref{eq:bnd2bndA}. Using Eq.~\eqref{eq:SvCrystal} and Eq.~\eqref{eq:bndACrystal} we obtain that the specimen GBND
\begin{equation}
  \label{eq:bnda2bnd}    
  \begin{split}
    \mathtt{BND}(\vec n)
    &= \int_{SO(3)} \mathtt{ODF}_{A}(g_{A}) 
    \int_{SO(3)} \mathtt{BCD}_A(\mathtt{inv}(g_{A}) g_B, \mathtt{inv}(g_A) \vec n) \d{g_B} \d{g_A}\\
    &= \int_{SO(3)} \mathtt{ODF}_{A}(g_{A}) 
    \int_{SO(3)} \mathtt{BCD}_A(\Delta g, \mathtt{inv}(g_A) \vec n) \d{\Delta g} \d{g_A}
    = \mathtt{ODF}_{A} \ast \mathtt{BND}_A (\vec n) 
  \end{split}
\end{equation}
is given by the spherical convolution of the ODF of the $A$-grains with the
crystal GBND of the $A$-grains. In other words in  crystallographically driven boundary networks the ODF and the crystal GBND determine the specimen GBND.

In summary, Eq.~\eqref{eq:bnda2bnd} leads to the following conclusions for the
specimen GBND of a crystallographically driven boundary network:
\begin{enumerate}
\item In an untextured material, i.e.\ $\mathtt{ODF}_{A} = 1$, the specimen
  GBND is uniform, i.e.\ $\mathtt{BND} = 1$.
\item In a textured material, i.e.\ $\mathtt{ODF}_{A} \ne 1$, a non-uniform
  crystal GBND, $\mathtt{BND}_{A} \ne 1$, may result in a non-uniform
  specimen GBND, $\mathtt{BND} \ne 1$.
\item Since convolution acts as a smoothing operation, the specimen GBND is
  typically less sharp than the crystal GBND.
\end{enumerate}

Such implications of the crystallographically driven boundary networks resulting in preferred crystal shapes are frequently found in materials science and geoscience, e.g.
\begin{itemize}
    \item In martensitic phase transformations of an untextured parent phase, the
  child grains typically exhibit no preferred alignment with respect to the
  specimen reference frame, consistent with (1).

    \item When deformation twins form in textured material, the alignment of twin boundary planes within a grain relates to the local crystallography of these grains and the global alignment depends on the orientation of the crystal in the entire material. Similarly, aggregates formed by alignment of euhedral platy grains e.g. due to deformation or settling would be representative of case (2).
    
    \item Similarly, euhedral crystals grown in veins where growth competition is dominant or along a gradient exhibit crystallographically controlled shapes but also an alignment with respect to the sample due to their growth history. 
\end{itemize}

It is noteworthy that Eq.~\eqref{eq:bnda2bnd} expresses the specimen GBND
in terms of the ODF and the crystal GBND restricted to the $A$-grains. This
is not restrictive as long as one can experimentally distinguish between $A$
and $B$ grains, or when both follow identical distributions, i.e.\
$\mathtt{ODF}_{A} = \mathtt{ODF}_{B}$ and
$\mathtt{BND}_A = \mathtt{BND}_{B}$. 

However, this becomes a limitation in situations where such a distinction is
not possible. For example, in twinning microstructures where host and twinned
grains cannot be reliably separated, but exhibit different orientation
distributions, the representation in Eq.~\eqref{eq:bnda2bnd} cannot be applied
directly.

\begin{figure}
  \begin{subfigure}{0.34\linewidth}
    \includegraphics[width=\linewidth]{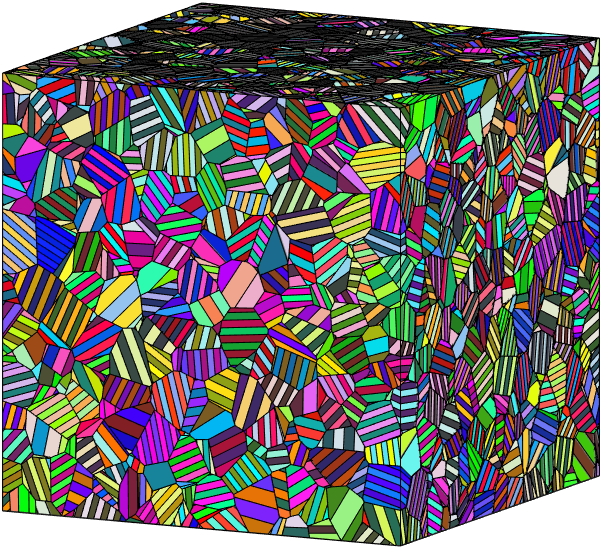}
    \caption{untextured microstructure}\label{fig:grainDynamicUniform}
  \end{subfigure}
  \hfill
  \begin{subfigure}{0.26\linewidth}
    \includegraphics[width=\linewidth]{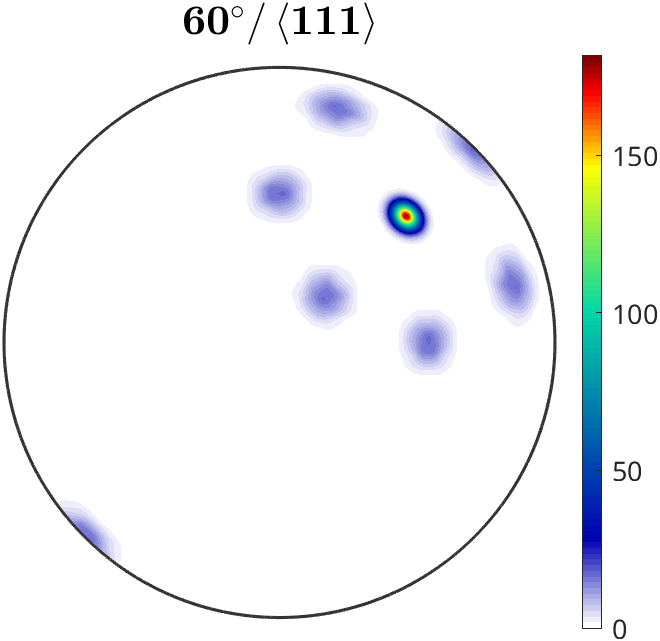}
    \caption{measured $\mathtt{BCD}_{AB}^{*}$}\label{fig:dynamicBCD}
  \end{subfigure}
  \hfill
  \begin{subfigure}{0.34\linewidth}
    \includegraphics[width=\linewidth]{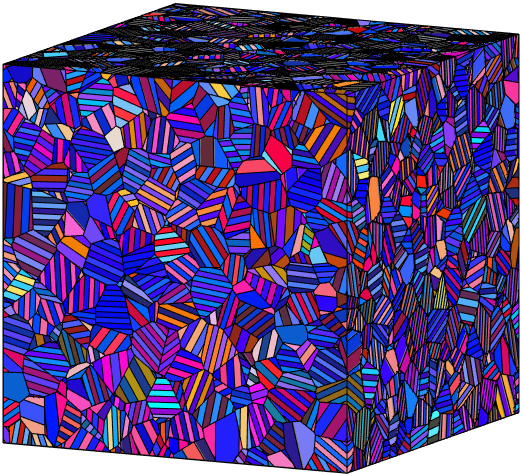}
    \caption{textured microstructure}\label{fig:grainDynamicTex}
  \end{subfigure}

  \bigskip
  
  \begin{subfigure}{0.24\linewidth}
    \includegraphics[width=\linewidth]{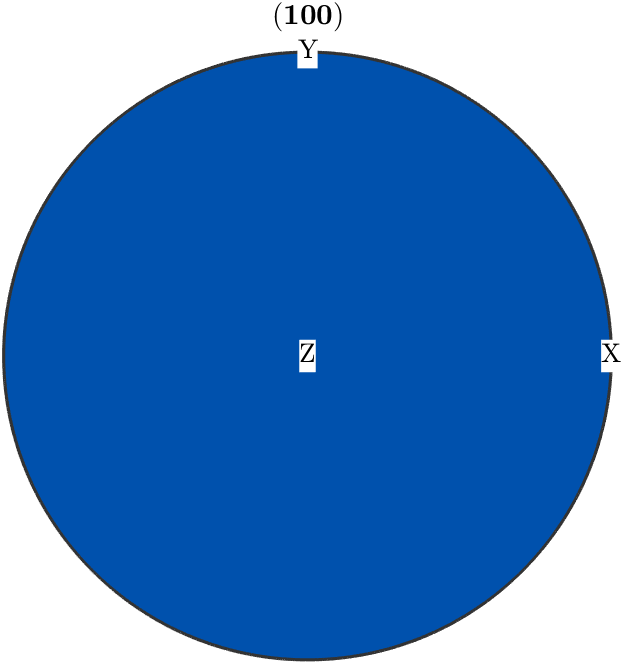}
    \caption{initial $\mathtt{ODF}$}\label{fig:pfDynamic}
  \end{subfigure}
  \begin{subfigure}{0.26\linewidth}
    \includegraphics[width=\linewidth]{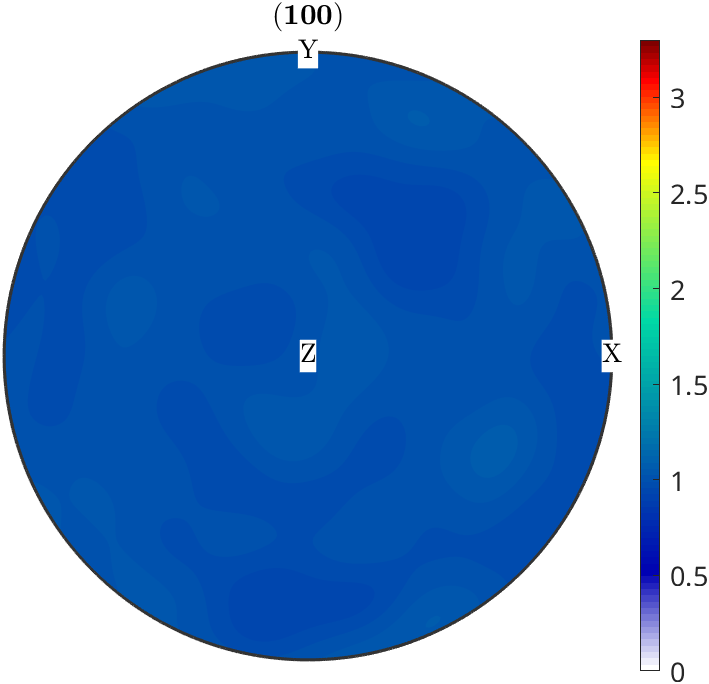}
    \caption{measured $\mathtt{ODF}$}\label{fig:pfDynamicSim}
  \end{subfigure}\hfill
  \begin{subfigure}{0.24\linewidth}
    \includegraphics[width=\linewidth]{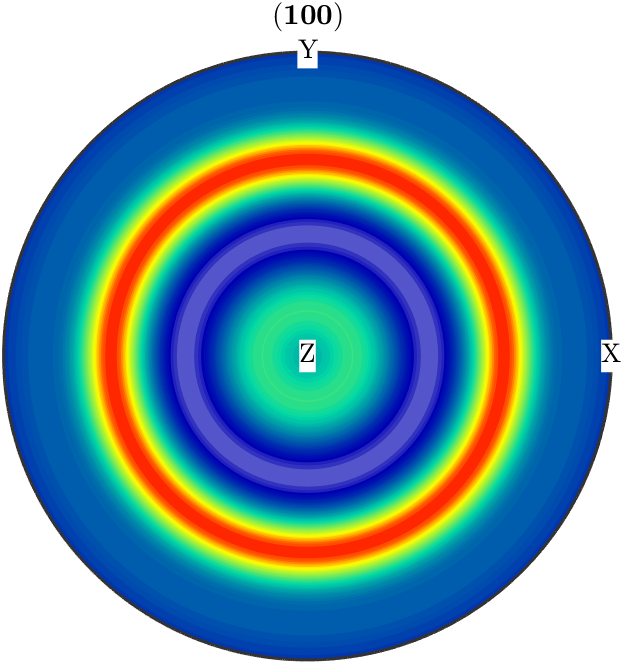}
    \caption{initial $\mathtt{ODF}$}\label{fig:pfDynamicTex}
  \end{subfigure}
  \begin{subfigure}{0.24\linewidth}
    \includegraphics[width=\linewidth]{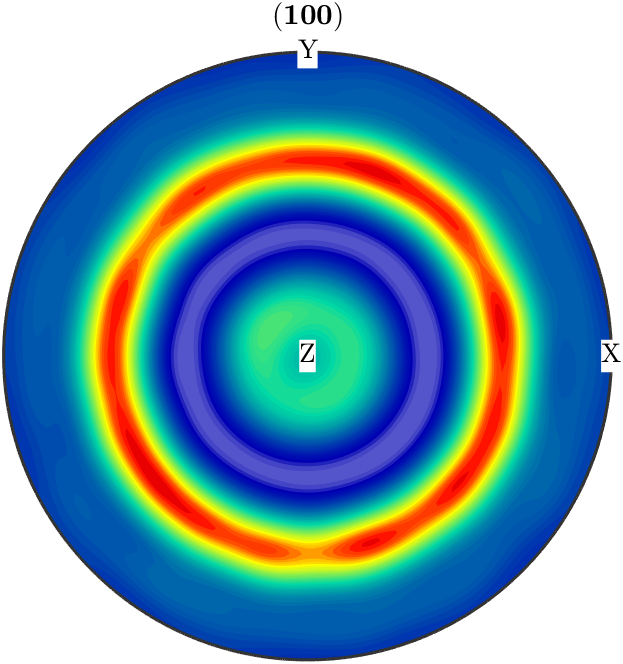}
    \caption{measured $\mathtt{ODF}$}\label{fig:pfDynamicTexSim}
  \end{subfigure}

  \bigskip
  
  \begin{subfigure}{0.24\linewidth}
    \includegraphics[width=\linewidth]{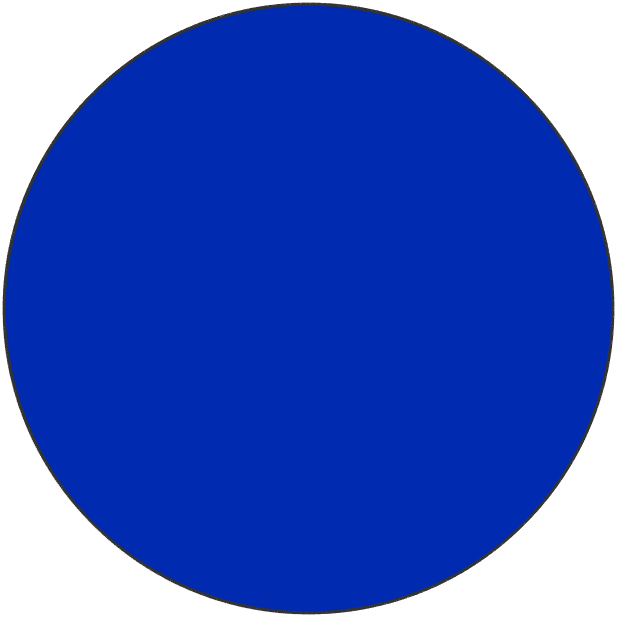}
    \caption{predicted $\mathtt{BND}$}\label{fig:bndDynamicUniform}
  \end{subfigure}
  \begin{subfigure}{0.26\linewidth}
    \includegraphics[width=\linewidth]{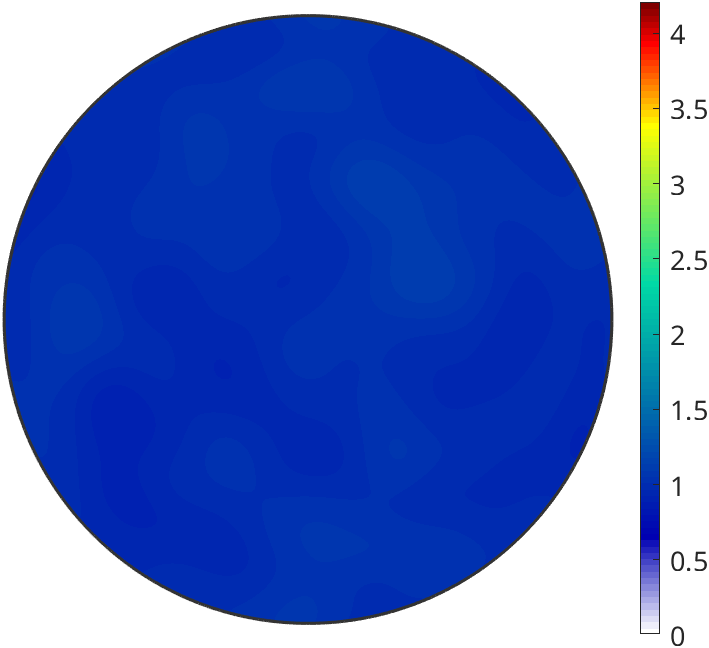}
    \caption{measured $\mathtt{BND}^{*}$}\label{fig:bndDynamicUniformSim}
  \end{subfigure}\hfill
  \begin{subfigure}{0.24\linewidth}
    \includegraphics[width=\linewidth]{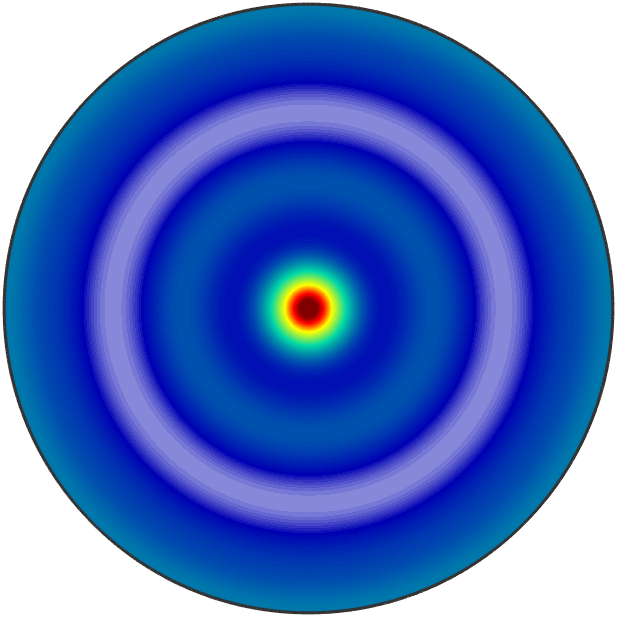}
    \caption{predicted $\mathtt{BND}$}\label{fig:bndDynamicTex}
  \end{subfigure}
  \begin{subfigure}{0.24\linewidth}
    \includegraphics[width=\linewidth]{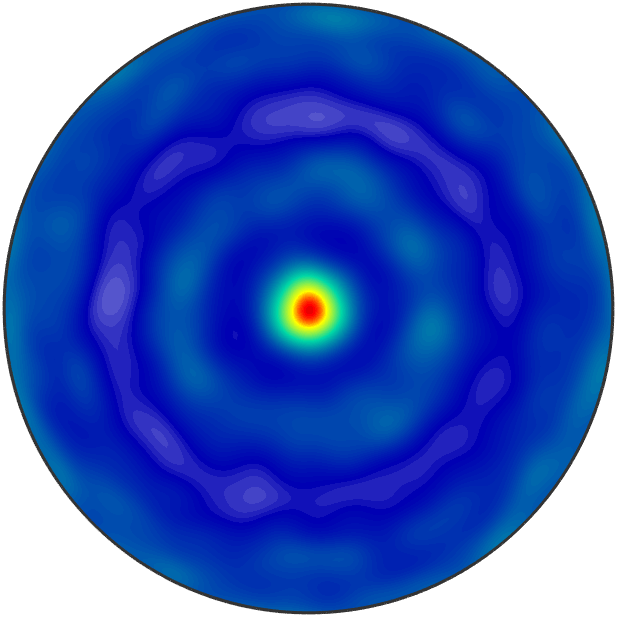}
    \caption{measured $\mathtt{BND}^{*}$}\label{fig:bndDynamicTexSim}
  \end{subfigure}
  
  \caption{Simulated cubic microstructure with $\Sigma3$ twinning resulting in
    the GBCD \subref{fig:dynamicBCD}. In \subref{fig:grainDynamicUniform} the
    grain orientations have been sampled from the uniform distribution
    $\mathtt{ODF}=1$ \subref{fig:pfDynamic} and in
    \subref{fig:grainDynamicTex} from the fibre ODF depicted in
    \subref{fig:pfDynamicTex}. The actual textures of the simulated
    microstructures are depicted in \subref{fig:pfDynamicSim} and
    \subref{fig:pfDynamicTexSim}, respectively. The resulting specimen GBNDs are
    depicted in \subref{fig:bndDynamicUniformSim} and
    \subref{fig:bndDynamicTexSim} for the untextured and textured
    microstructures, respectively. The theoretical predictions
    $\mathtt{BND} = \mathtt{ODF}_{A} \ast \mathtt{BND}_{A}$, obtained
    by Eq.~\eqref{eq:bnda2bnd} using the theoretical ODF
    and the theoretical crystal GBND.}\label{fig:Dynamic}
\end{figure}

\begin{figure}
    \centering
        
    \begin{subfigure}{0.175\linewidth}
    \includegraphics[width=\linewidth]{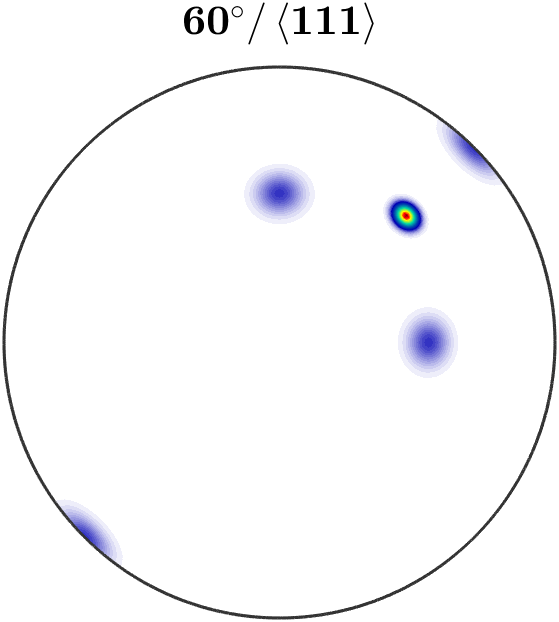}
    \caption{ $\mathtt{BCD}_A$} \label{fig:bndA}
  \end{subfigure}  
    \begin{subfigure}{0.20\linewidth}
    \includegraphics[width=\linewidth]{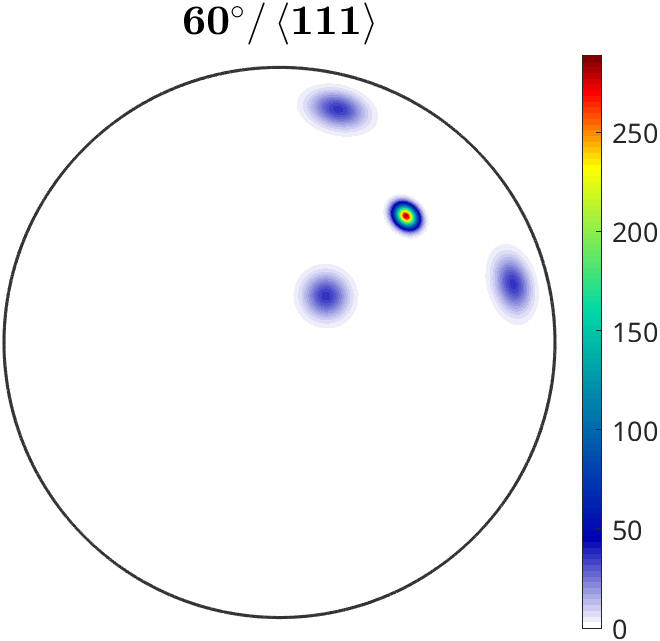}
    \caption{ $\mathtt{BCD}_{B}$}\label{fig:bndB}
  \end{subfigure}
    \begin{subfigure}{0.175\linewidth}
    \includegraphics[width=\linewidth]{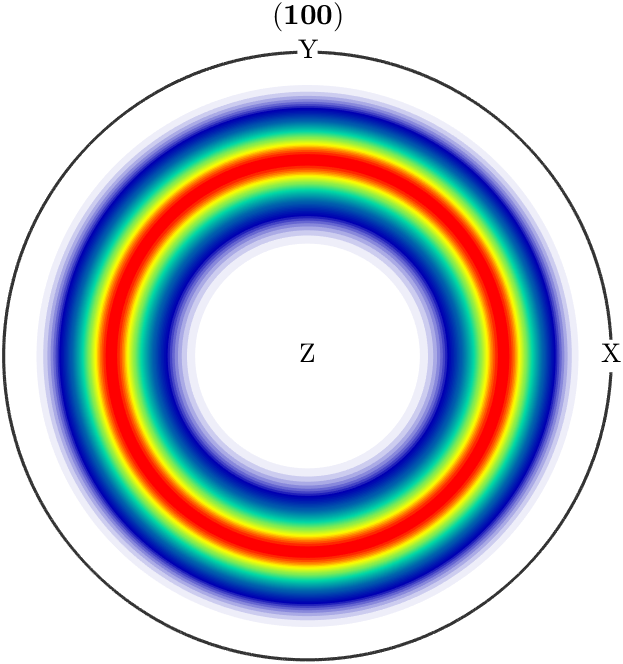}
    \caption{ $\mathtt{ODF}_{A}$}\label{fig:odfA}
  \end{subfigure} 
  \begin{subfigure}{0.195\linewidth}
    \includegraphics[width=\linewidth]{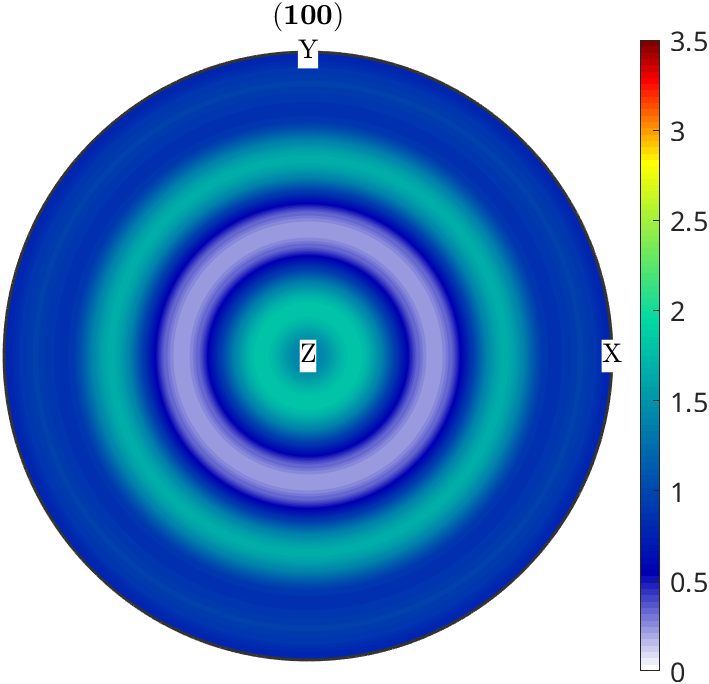}
    \caption{ $\mathtt{ODF}_{B}$}\label{fig:odfB}
  \end{subfigure}
  \begin{subfigure}{0.2\linewidth}
    \includegraphics[width=\linewidth]{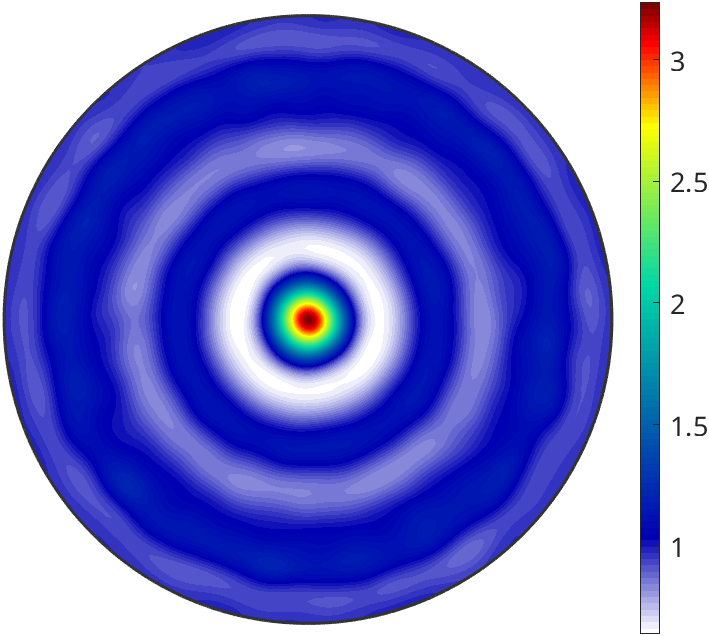}
    \caption{ predicted \texttt{BND}}\label{fig:gbndDynamicTexAB}
  \end{subfigure}
  
  \caption{The initial GBCD with respect to the host grains \subref{fig:bndA}
    and the twinned grains \subref{fig:bndB} used for the simulation of
    3d-microstructures in Fig.~\ref{fig:Dynamic}. The $(100)$ pole figures of
    the initial ODF for the host grains \subref{fig:odfA} and the resulting
    ODF predicted for the twinned grains \subref{fig:odfB} used in the
    simulation of the textured microstructure in
    Fig.~\ref{fig:grainDynamicTex}. The predicted specimen GBND in
    \subref{fig:gbndDynamicTexAB} was obtained as the convolution
    $\mathtt{ODF}^{*}_{AB} \ast \mathtt{BND}_{AB}^{*}$ of the
    measured ODF with the measured crystal GBND and should be compared with the
    more accurate prediction
    $\mathtt{BND} = \mathtt{ODF}_{A} \ast \mathtt{BND}_{A}$ depicted in
    Fig.~\ref{fig:bndDynamicTex}.}
\end{figure}

\paragraph{Simulated example of a crystallographically driven microstructure}
In order to illustrate the above findings we simulate an untextured three-dimensional FCC microstructure consisting of about 33\,000 grains, i.e. $\mathtt{ODF}=1$, all containing $\Sigma 3$ twins, cf. Fig.~\ref{fig:grainDynamicUniform}. The habit planes of those twins are simulated according to the GBCD as depicted in  Fig.~\ref{fig:bndA}.

We have purposely chosen the GBCD in a way that
$\mathtt{BCD}_A \ne \mathtt{BCD}_{B}$. The difference between those two
distribution functions is depicted in Fig.~\ref{fig:bndA} and
Fig.~\ref{fig:bndB}. Without additional information we cannot distinguish
between host and twinned grains in our simulated microstructure. Hence, we are
in the setting of grain exchange symmetry. The GBCD measured from the
3d-microstructure, Fig.~\ref{fig:dynamicBCD}, shows indeed a very good fit to
the average $\tfrac12(\texttt{BCD}_{A} + \texttt{BCD}_{B})$, indicating that the
simulation procedure is accurate.

Fig.~\ref{fig:pfDynamicSim} displays the $(100)$ pole figure of the ODF estimated from the simulated 3d-microstructure, confirming a uniform texture. According to our theory this should lead to a uniform distribution of the boundary normal in specimen coordinates, i.e. $\mathtt{BND}=1$. Computing the specimen GBND from the simulated microstructure by kernel density estimation we indeed observe an almost uniform distribution as depicted in Fig.~\ref{fig:bndDynamicUniformSim}.

For the second experiment we keep the twinning relationship and the GBCD, but choose the orientations of the host ($A$) grains to form a $\gamma$-fiber texture, cf. Fig.~\ref{fig:odfA}. This implies a texture of the twinned $B$ grains as shown in Fig.~\ref{fig:odfB}. The resulting microstructure together with the $(100)$ pole figure of twinned and host grains are depicted in Fig.~\ref{fig:grainDynamicTex} and \ref{fig:pfDynamicTexSim}.

We clearly observe that the specimen GBND computed from the textured
3d-microstructure, Fig.~\ref{fig:bndDynamicTexSim}, is not uniform anymore,
but shows a fiber structure with a clear maximum in $Z$ direction. Comparing the specimen GBND, Fig.~\ref{fig:bndDynamicTexSim}, measured from the
3d-microstructure, with the theoretical prediction in
Fig.~\ref{fig:bndDynamicTex} obtained by  
$\mathtt{BND} = \mathtt{ODF}_{A} \ast \mathtt{BND}_{A}$ according to
Eq.~\eqref{eq:bnda2bnd}, we observe a very good fit. As already discussed
above, in practical applications we often can only measure the joined
orientations distribution function $\mathtt{ODF}^{*}$, Fig.~\ref{fig:pfDynamicTexSim}, and the crystal
GBND $\mathtt{BND}_{AB}^{*}$ modulo grain exchange symmetry. Plugging these
measured distributions into the above equation, i.e., $\mathtt{BND} \approx
\mathtt{ODF}^{*} \ast \mathtt{BND}_{AB}^{*}$, we arrive at the distribution
depicted in Fig.~\ref{fig:gbndDynamicTexAB}, which is still quite similar to
the measured specimen GBND in Fig.~\ref{fig:bndDynamicTexSim}.

\subsection{Mixed Boundary Networks}
\label{sec:mixed-bound-netw}

In the previous sections we considered two limiting cases of grain boundary
networks, namely macroscopically driven and crystallographically driven
boundary distributions. In practice, however, microstructures are often
influenced by a combination of both mechanisms.

\begin{figure}
  \begin{subfigure}{0.34\linewidth}
    \includegraphics[width=\linewidth]{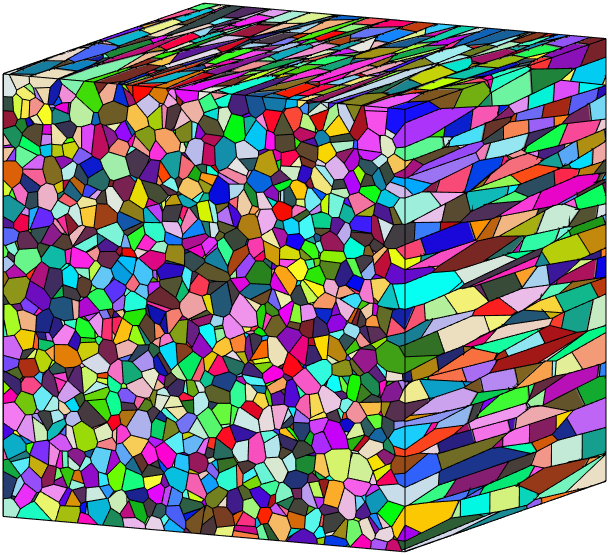}
    \caption{untextured microstructure}\label{fig:grainMixUniform}
  \end{subfigure}
  \hfill
  \begin{subfigure}{0.26\linewidth}    
    \includegraphics[width=\linewidth]{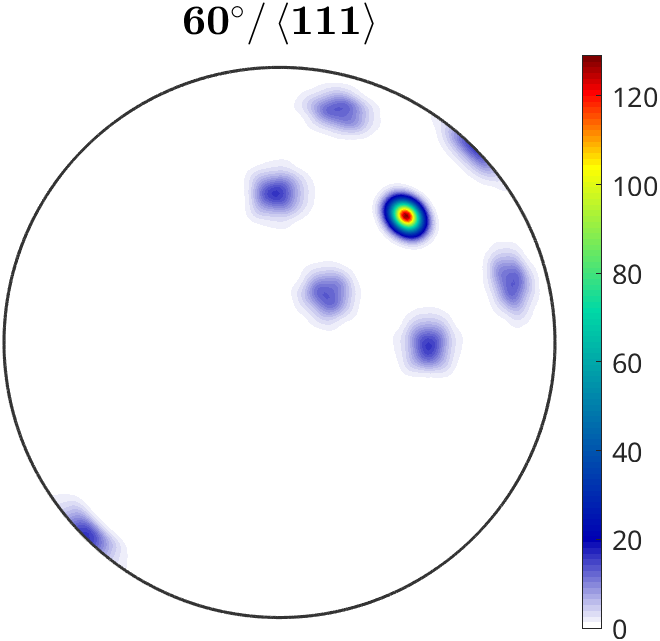}\\
    \includegraphics[width=\linewidth]{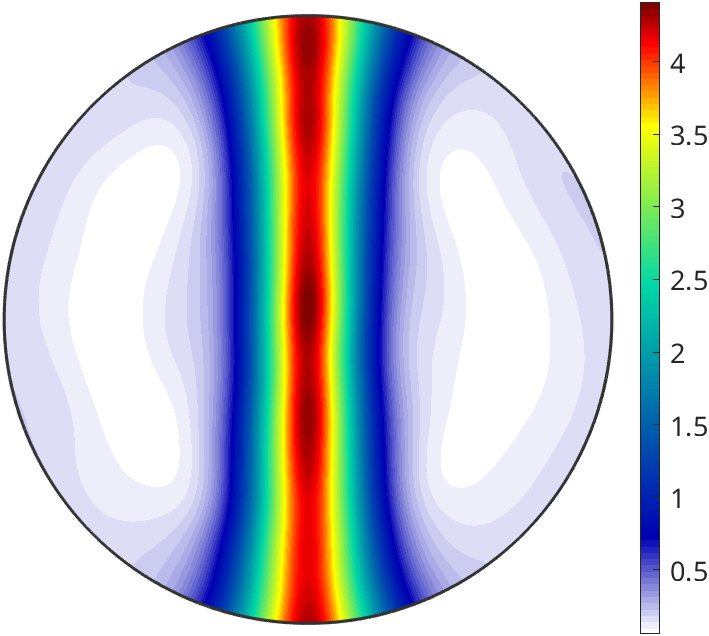}
    \caption{$\mathtt{GBCD}$ and $\mathtt{GBND}$}\label{fig:MixedBCD}
  \end{subfigure}
  \hfill
  \begin{subfigure}{0.34\linewidth}
    \includegraphics[width=\linewidth]{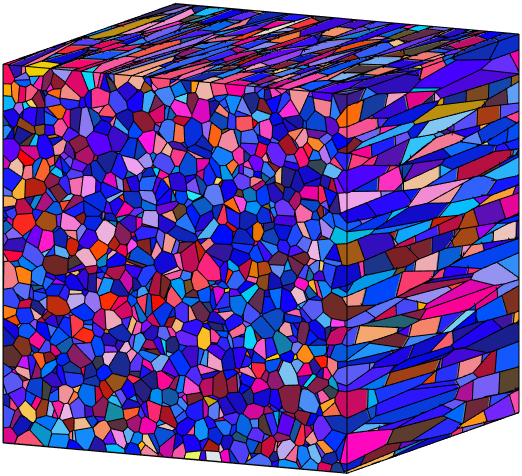}
    \caption{textured microstructure}\label{fig:grainMixTex}
  \end{subfigure}

  \bigskip
  
  \begin{subfigure}{0.24\linewidth}
    \includegraphics[width=\linewidth]{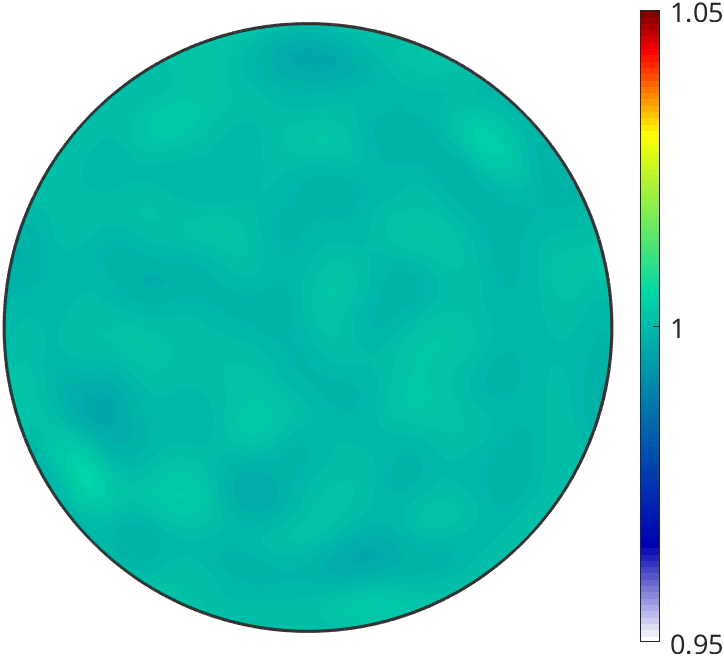}
    \caption{predicted $\mathtt{BND}$}
    \label{fig:gbndMixUniform}
  \end{subfigure}
  \begin{subfigure}{0.26\linewidth}
    \includegraphics[width=\linewidth]{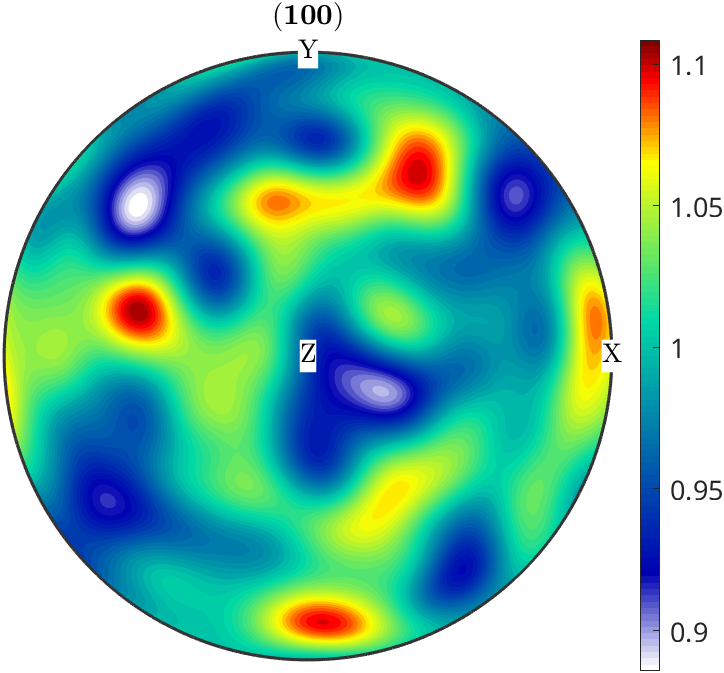}
    \caption{measured $\mathtt{ODF}$}
    \label{fig:pfMixUniformSim}
  \end{subfigure}\hfill
  \begin{subfigure}{0.24\linewidth}
    \includegraphics[width=\linewidth]{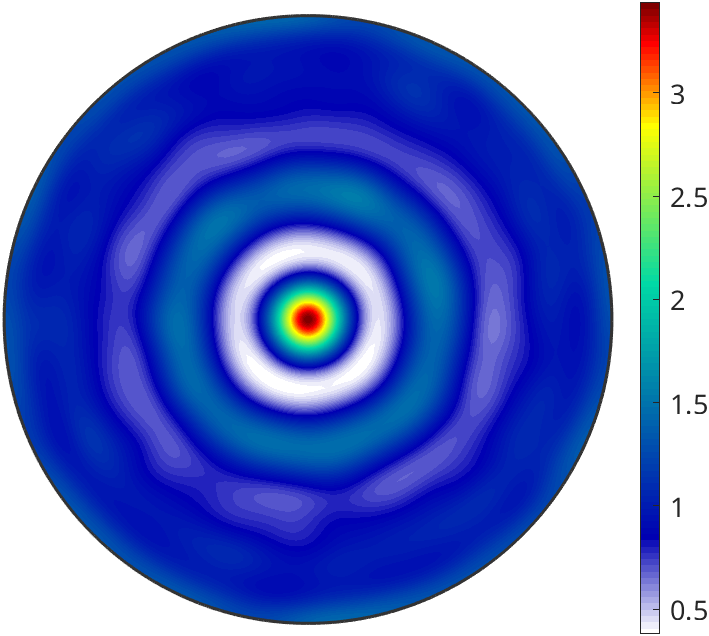}
    \caption{predicted $\mathtt{BND}$}
    \label{fig:gbndMixTex}
  \end{subfigure}
  \begin{subfigure}{0.24\linewidth}
    \includegraphics[width=\linewidth]{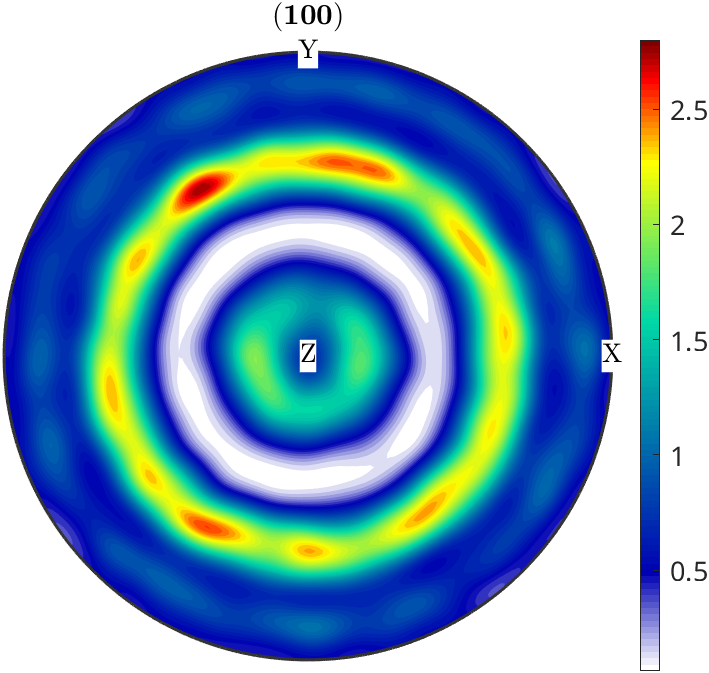}
    \caption{measured $\mathtt{ODF}$}\label{fig:pfMixTexSim}
  \end{subfigure}

  \bigskip
  
  \begin{subfigure}{0.245\linewidth}
    \includegraphics[width=\linewidth]{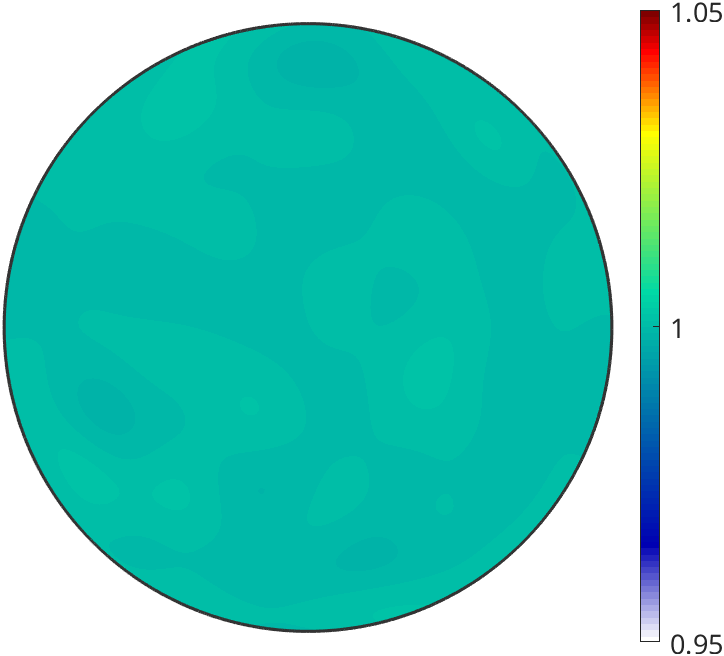}
    \caption{predicted $\mathtt{BND}_A$}
    \label{fig:gbndABMixUniform}
  \end{subfigure}
  \begin{subfigure}{0.2456\linewidth}
    \includegraphics[width=\linewidth]{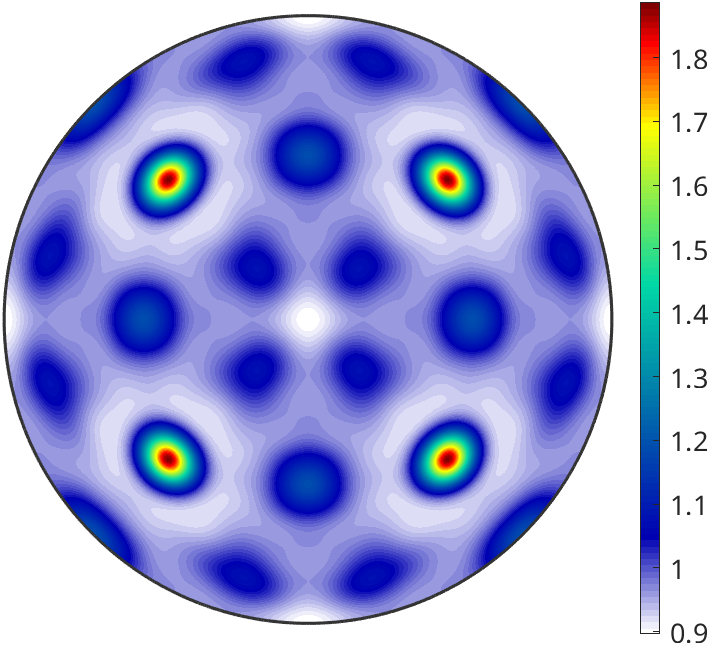}
    \caption{measured $\mathtt{BND}_A$}
    \label{fig:gbndABMixUniformSim}
  \end{subfigure}\hfill
  \begin{subfigure}{0.245\linewidth}
    \includegraphics[width=\linewidth]{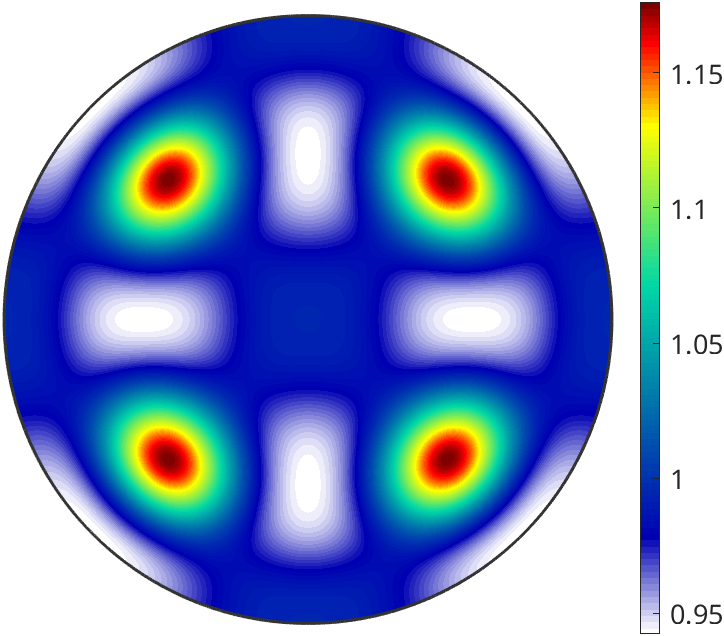}
    \caption{predicted $\mathtt{BND}_A$}
    \label{fig:gbndABMixTex}
  \end{subfigure}
  \begin{subfigure}{0.245\linewidth}
    \includegraphics[width=\linewidth]{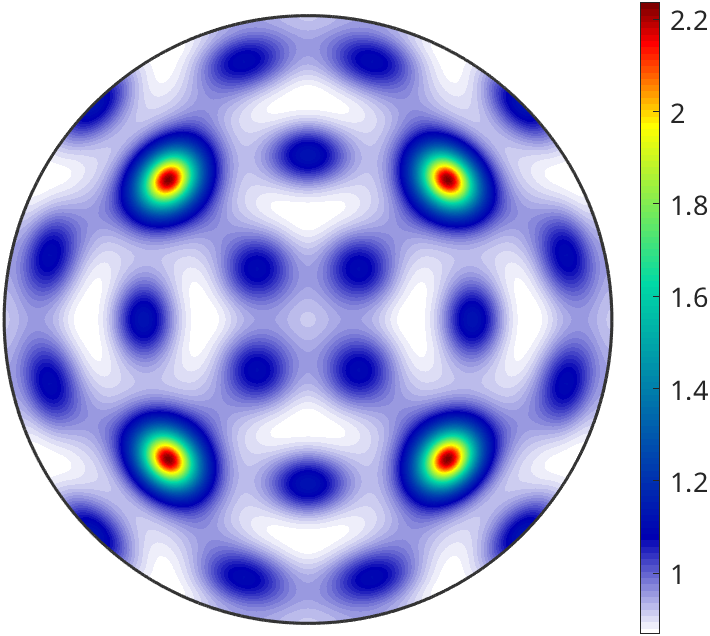}
    \caption{measured $\mathtt{BND}_{A}$}
    \label{fig:gbndABMixTexSim}
  \end{subfigure}
  
  \caption{Simulated cubic microstructures with a preferred grain elongation
    in x-direction and $\Sigma3$ twinning resulting in the specimen GBND and
    the GBCD as depicted in \subref{fig:MixedBCD}. In
    \subref{fig:grainMixUniform} the grain orientations are distributed
    uniformly as depicted in $(100)$ pole figure \subref{fig:pfMixUniformSim},
    whereas in \subref{fig:grainMixTex} the grain orientations form a fiber
    texture \subref{fig:pfMixTexSim}.
    The theoretical predictions of the specimen GBNDs assuming a
    crystallographically driven grain boundary network are depicted in
    \subref{fig:gbndMixUniform} and \subref{fig:gbndMixTex}.
    The theoretical predictions of the crystal GBNDs assuming a
    macroscopically driven grain boundary network are depicted in
    \subref{fig:gbndABMixUniform} and \subref{fig:gbndABMixTex}. These should
    be compared to the corresponding measured crystal GBNDs in
    \subref{fig:gbndABMixUniformSim} and \subref{fig:gbndABMixTexSim}.}
  \label{fig:Mixed}
\end{figure}

To illustrate such a mixed situation, we consider the macroscopically driven
grain structure from Section \ref{sec:macr-init-bound} with either uniformly
distributed grain orientations, Fig.~\ref{fig:grainMixUniform} or grain
orientations derived from a $\gamma$-fibre ODF, Fig.~\ref{fig:grainMixTex}. In a
second step each grain is subdivided into $\Sigma3$ twins with interface
planes distributed according to the distribution in Fig.~\ref{fig:dynamicBCD}.
This leads to a boundary network in which both macroscopic alignment and
crystallographic selection contribute to the observed distributions. Note that by construction, both the untextured and the textured microstructures share the same specimen GBND, crystal GBND and GBCD, Figs.~\ref{fig:MixedBCD} and \ref{fig:gbndABMixTex}.

We first consider the hypothesis of a purely macroscopically driven boundary
network. In this case the measured specimen GBND is prescribed and the model
\eqref{eq:svMacro} implies a prediction
$\mathtt{ODF}^{*} \circledast \mathtt{BND}^{*}$ for the corresponding crystal
GBND as the spherical convolution of the measured ODF with the measured
specimen GBND. Not surprisingly, the predicted crystal GBND,
Fig.~\ref{fig:gbndABMixUniform}, for the untextured case is almost completely
uniform. In the textured case the predicted crystal GBND,
Fig.~\ref{fig:gbndABMixTex}, shows some similarity with the measured GBND,
Fig.~\ref{fig:gbndABMixTexSim}. However, in both scenarios the prediction
exhibits substantially lower directional contrast than the measurement. In a practical application this would indicate that the hypothesis of a purely macroscopically driven boundary network is not sufficient to explain the observations.

Conversely, we consider the hypothesis of a purely crystallographically driven
boundary network. In this case, the crystal GBND is prescribed and the model
\eqref{eq:SvCrystal} implies a prediction
$\mathtt{ODF}^{*} \ast \mathtt{BND}^{*}_{AB}$ for the corresponding specimen
GBND, obtained as the spherical convolution of the measured ODF with the
measured crystal GBND. For the untextured microstructure, the predicted
specimen GBND, Fig.~\ref{fig:gbndMixUniform}, is again nearly uniform. In the
textured case, however, the predicted specimen GBND,
Fig.~\ref{fig:gbndMixTex}, is strongly anisotropic but differs significantly
from the measured specimen GBND shown in Fig.~\ref{fig:MixedBCD}. This
indicates that the hypothesis of a purely crystallographically driven boundary
network is also not sufficient.

Taken together, these results show that neither of the two limiting models is
sufficient to explain the observed distributions. The boundary network must
therefore be described as a combination of macroscopic alignment and
crystallographic selection effects.

\section{Discussion}

The results derived in this work highlight a fundamental ambiguity in the
interpretation of grain boundary plane distributions. In particular, we have
shown that preferred boundary plane orientations observed in a given reference
frame do not uniquely determine the underlying mechanism of boundary formation.

For macroscopically driven boundary networks, the crystal GBND is given by a
convolution of the specimen GBND with the orientation distribution function
(ODF). Conversely, for crystallographically driven boundary networks, the
specimen GBND is obtained as a convolution of the crystal GBND with the
ODF. This duality implies that anisotropies observed in one reference frame
may arise either from intrinsic crystallographic selection or from macroscopic
alignment effects combined with texture.

A direct consequence of this result is that a non-uniform crystal-frame GBND
in a textured material does not necessarily imply crystallographic control of
grain boundary formation. Instead, such anisotropies may arise purely from
macroscopic processes that align grain shapes, such as deformation, growth, or
recrystallization, in combination with a non-random orientation distribution.
Similarly, anisotropies observed in the specimen reference frame may reflect
underlying crystallographic mechanisms, even in the absence of macroscopic
alignment.

From a mathematical perspective, the convolution relations derived in this work
show that the ODF acts as an angular averaging operator on the GBND. As a
result, directional features of the boundary normal distribution are typically
smoothed when transformed between reference frames. The extent of this
smoothing depends on the sharpness of the ODF and may obscure or attenuate
features associated with the underlying boundary formation mechanism.

In practice, real microstructures are unlikely to fall into either of the two
limiting cases considered here. Instead, boundary networks are typically
influenced by a combination of macroscopic and crystallographic effects. In
such cases, the observed GBND may reflect a superposition of both mechanisms,
and the convolution relations derived above provide a useful framework for
understanding how these contributions interact.

These findings have important implications for the interpretation of
experimentally measured grain boundary distributions. In particular, care must
be taken when attributing preferred boundary plane orientations to specific
mechanisms based solely on GBND measurements in a single reference frame.
Additional information, such as the ODF, misorientation distributions, or
independent knowledge of the processing history, is required to disentangle
macroscopic and crystallographic contributions.

Finally, the framework proposed here suggests a possible route for
distinguishing between different boundary formation mechanisms. By comparing
GBNDs in specimen and crystal reference frames and accounting for the effect
of the ODF, it may be possible to identify signatures that are characteristic
of macroscopic alignment or crystallographic selection. Developing robust
statistical tests for this purpose remains an interesting direction for future
work.



\bibliographystyle{elsarticle-num}   
\bibliography{literatur}

@book{Morawiec2004Book,
  author = {Morawiec, A.},
  title  = {Orientations and Rotations: Computations in Crystallographic Textures},
  publisher = {Springer},
  year   = {2004}
}

@book{GottsteinShvindlerman2010GBMigration,
  author = {Gottstein, G. and Shvindlerman, L. S.},
  title  = {Grain Boundary Migration in Metals: Thermodynamics, Kinetics, Applications},
  publisher = {CRC Press},
  year   = {2010}
}

@Article{MTEX,
  author = 	 {F. Bachmann and R. Hielscher and H. Schaeben},
  title = 	 {Texture Analysis with MTEX - Free and Open Source Software Toolbox},
  journal = 	 {Solid State Phenomena},
  year = 	 2010,
  volume = 	 169,
  pages = 	 {63-68}}

@article{Herring1951SurfaceEnergy,
  author  = {Herring, C.},
  title   = {Some Theorems on the Free Energies of Crystal Surfaces},
  journal = {Physical Review},
  year    = {1951},
  volume  = {82},
  pages   = {87--93}
}

@article{ChristianMahajan1995Twins,
  author  = {Christian, J. W. and Mahajan, S.},
  title   = {Deformation Twinning},
  journal = {Progress in Materials Science},
  year    = {1995},
  volume  = {39},
  pages   = {1--157},
}

@article{Cahn1964Interfaces,
  author  = {Cahn, J. W.},
  title   = {On the Morphology of Grain and Phase Boundaries},
  journal = {Acta Metallurgica},
  year    = {1964},
  volume  = {12},
  pages   = {1183--1184}
}

@book{PorterEasterling2009PhaseTransformations,
  author = {Porter, D. A. and Easterling, K. E. and Sherif, M. Y.},
  title  = {Phase Transformations in Metals and Alloys},
  edition = {3},
  publisher = {CRC Press},
  year   = {2009}
}

@article{Wulff1901CrystalShape,
  author  = {Wulff, G.},
  title   = {Zur Frage der Geschwindigkeit des Wachstums und der Auflösung der Kristallflächen},
  journal = {Zeitschrift für Kristallographie},
  year    = {1901},
  volume  = {34},
  pages   = {449--530}
}

@article{Rutter1976PressureSolution,
  author  = {Rutter, E. H.},
  title   = {The Kinetics of Rock Deformation by Pressure Solution},
  journal = {Philosophical Transactions of the Royal Society A},
  year    = {1976},
  volume  = {283},
  pages   = {203--219},
}

@article{LIFSHITZ1963,
  author  = {Lifshitz, I.M.},
  title   = {On the theory of diffusion-viscous flow of polycrystalline bodies},
  journal = {Soviet Phys. JETP},
  year    = {1963},
  volume  = {17},
  pages   = {909--920},
  doi     = {},
}

@article{FordWheelerMovchan2002,
  author  = {Ford, J.M. and Wheeler, J. and Movchan, A.B.},
  title   = {Computer simulation of grain-boundary diffusion creep},
  journal = {Acta Materialia},
  year    = {2002},
  volume  = {50},
  pages   = {3941--3955},
}

@article{Chin1973,
  author  = {Chin, G.Y.},
  title   = {A Theoretical examination of the plastic deformation of ionic crystals: I. maximum work analysis for slip on {110} 〈110〉 and {100} 〈110〉 systems.},
  journal = {METALLURGICALTRANSACTIONS},
  year    = {1973},
  volume  = {4},
  pages   = {329--333},
}

@article{URAI1991,
author = {Urai, J.L. and Williams, P.F. and {van Roermund}, H.L.M. },
title = {Kinematics of crystal growth in syntectonic fibrous veins},
journal = {Journal of Structural Geology},
volume = {13},
number = {7},
year = {1991},
pages = {823-836},
}

@article{MEANS2001,
title = {A laboratory simulation of fibrous veins: some first observations},
author = {Means, W.D. and Li, T.},
journal = {Journal of Structural Geology},
volume = {23},
number = {6},
pages = {857-863},
year = {2001},
issn = {0191-8141},
}

@article{Bons1997Exp,
  title={Experimental simulation of the formation fibrous veins by localised dissolution-precipitation creep},
  author={Bons, P.D. and Jessell, M.W.},
  journal={Mineralogical Magazine},
  year={1997},
  volume={61},
  pages={53 - 63},
}

@article{Rohrer2011,
  author  = {Rohrer, G. S.},
  title   = {Grain Boundary Energy Anisotropy: A Review},
  journal = {Journal of Materials Science},
  year    = {2011},
  volume  = {46},
  number  = {18},
  pages   = {5881--5895},
}

@article{Rollett2007,
   author = "Rollett, Anthony D. and Lee, S.-B. and Campman, R. and Rohrer, G.S.",
   title = "Three-Dimensional Characterization of Microstructure by Electron Back-Scatter Diffraction", 
   journal= "Annual Review of Materials Research",
   year = "2007",
   volume = "37",
   number = "Volume 37, 2007",
   pages = "627-658",
   publisher = "Annual Reviews",
   issn = "1545-4118",
   type = "Journal Article",
}

@article{QUEY20111729,
title = {Large-scale 3D random polycrystals for the finite element method: Generation, meshing and remeshing},
journal = {Computer Methods in Applied Mechanics and Engineering},
volume = {200},
number = {17},
pages = {1729-1745},
year = {2011},
issn = {0045-7825},
author = {R. Quey and P.R. Dawson and F. Barbe},
}

@article{quey:hal-01626440,
  TITLE = {{Optimal polyhedral description of 3D polycrystals: method and application to statistical and synchrotron X-ray diffraction data}},
  AUTHOR = {Quey, Romain and Renversade, Lo{\"i}c},
  JOURNAL = {{Computer Methods in Applied Mechanics and Engineering}},
  PUBLISHER = {{Elsevier}},
  VOLUME = {330},
  PAGES = {308-333},
  YEAR = {2017},
  MONTH = Oct,
 }

@article{REE1991,
title = {An experimental steady-state foliation},
author = {Ree, J.H.},
journal = {Journal of Structural Geology},
volume = {13},
number = {9},
pages = {1001--1011},
year = {1991},
issn = {0191-8141},
}

@article{SCHMID1987,
title = {Simple shear experiments on calcite rocks: rheology and microfabric},
author = {Schmid, S.M. and Panozzo, R. and Bauer, S.},
journal = {Journal of Structural Geology},
volume = {9},
number = {5},
pages = {747-778},
year = {1987},
issn = {0191--8141},
}

@article{Haase2013,
title = {On the Relation of Microstructure and Texture Evolution in an Austenitic Fe-28Mn-0.28C TWIP Steel During Cold Rolling},
author = {Haase, C. and Chowdhury, S.G. and Barrales-Mora, L.A. and Molodov, D.A. and Gottstein, G.},
journal = {Metallurgical and Materials Transactions A},
volume = {2},
number = {44},
pages = {911--922},
year = {2013},
}

@article{Ferreira2021,
  title        = {The Effect of Grain Boundaries on Plastic Deformation of Olivine},
  author       = {Ferreira, Filippe and Hansen, Lars N. and Marquardt, Katharina},
  journal      = {Journal of Geophysical Research: Solid Earth},
  volume       = {126},
  number       = {7},
  year         = {2021},
  
}

@incollection{Rohrer2014,
  author       = {Gregory S. Rohrer},
  title        = {Microstructural Characterization of Hard Ceramics},
  booktitle    = {Comprehensive Hard Materials},
  editor       = {V. K. Sarin},
  publisher    = {Elsevier},
  year         = {2014},
  volume       = {2},
  pages        = {265--282}, 
}

@article{Marquardt2015,
  author       = {Katharina Marquardt and Gregory S. Rohrer and Luiz Morales and Erik Rybacki and Hauke Marquardt and Brian Lin},
  title        = {The most frequent interfaces in olivine aggregates: the GBCD and its importance for grain boundary related processes},
  journal      = {Contributions to Mineralogy and Petrology},
  volume       = {170},
  number       = {40},
  pages        = {1--17},
  year         = {2015},
  publisher    = {Springer-Verlag}
}

@article{Kohler2022,
  author       = {Felix Kohler and Olivier Pierre-Louis and Dag Kristian Dysthe},
  title        = {Crystal growth in confinement},
  journal      = {Nature Communications},
  volume       = {13},
  number       = {1},
  pages        = {6990},
  year         = {2022},
  publisher    = {Nature Publishing Group}
}

@article{Spear2024,
  author       = {Frank S. Spear},
  title        = {A grain boundary model of metamorphic reaction},
  journal      = {Contributions to Mineralogy and Petrology},
  volume       = {179},
  number       = {4},
  year         = {2024},
   publisher    = {Springer Nature},
}

@article{Wieser2020,
  author       = {Philippa E. Wieser and Marie Edmonds and John Maclennan and John Wheeler},
  title        = {Microstructural constraints on magmatic mushes under Kilauea Volcano, Hawai'i},
  journal      = {Nature Communications},
  volume       = {11},
  number       = {1},
  pages        = {14},
  year         = {2020},
   publisher    = {Nature Publishing Group},
}

@article{Wieser2019,
    author = {Wieser, P.E. and Vukmanovic, Z. and Kilian, R. and Ringe, E. and Holness, M.B. and Maclennan, J. and Edmonds, M.},
    title = {To sink, swim, twin, or nucleate: A critical appraisal of crystal aggregation processes},
    journal = {Geology},
    volume = {47},
    number = {10},
    pages = {948-952},
    year = {2019},
    month = {08},
}

@article{Rehn2019,
  author       = {Veronika Rehn and Johannes Hotzer and Wolfgang Rheinheimer and Marco Seiz and Christopher Serr and Britta Nestler},
  title        = {Phase-field study of grain growth in porous polycrystals},
  journal      = {Acta Materialia},
  volume       = {174},
  pages        = {439--449},
  year         = {2019},
  publisher    = {Elsevier}
}

@article{Austin2025,
  title        = {Grain boundary complexion transitions in olivine with temperature},
  author       = {Alexandra C. Austin and Sanae Koizumi and Martin Folwarczny and David P. Dobson and Katharina Marquardt},
  journal      = {arXiv e-print},
  volume       = {arXiv:2504.19784},
  year         = 2025,
  eprint       = {2504.19784},
  archivePrefix = {arXiv},
  primaryClass = {physics.geo-ph}
}

@article{Saylor2004PlanarSections,
  author  = {Saylor, David M. and El-Dasher, Bassem S. and Adams, Brent L. and Rohrer, Gregory S.},
  title   = {Measuring the five-parameter grain-boundary distribution from observations of planar sections},
  journal = {Metallurgical and Materials Transactions A},
  year    = {2004},
  volume  = {35},
  number  = {7},
  pages   = {1981--1989}, 
}

@article{LarsenAdams2004NewStereology,
  author  = {Larsen, Ryan J. and Adams, Brent L.},
  title   = {New stereology for the recovery of grain-boundary plane distributions in the crystal frame},
  journal = {Metallurgical and Materials Transactions A},
  year    = {2004},
  volume  = {35},
  number  = {7},
  pages   = {1991--1998},
}

@article{HallWatsonCabrera1987,
  author  = {Hall, Peter and Watson, Geoffrey S. and Cabrera, Javier},
  title   = {Kernel density estimation with spherical data},
  journal = {Biometrika},
  year    = {1987},
  volume  = {74},
  number  = {4},
  pages   = {751--762},
}

@book{Silverman1986,
  author    = {B. W. Silverman},
  title     = {Density Estimation for Statistics and Data Analysis},
  year      = {1986},
  publisher = {Chapman and Hall},
  address   = {London},
  series    = {Monographs on Statistics and Applied Probability},
  isbn      = {978-0412246203}
}

@article{AdamsField1992_GBTexture,
  author  = {B. L. Adams and D. P. Field},
  title   = {Measurement and representation of grain-boundary texture},
  journal = {Metallurgical and Materials Transactions A},
  year    = {1992},
  volume  = {23},
  pages   = {2501--2513},
}

@article{SaylorMorawiecRohrer2003_MgO5DOF,
  author  = {D. M. Saylor and A. Morawiec and G. S. Rohrer},
  title   = {Distribution of grain boundaries in magnesia as a function of five macroscopic parameters},
  journal = {Acta Materialia},
  year    = {2003},
  volume  = {51},
  number  = {13},
  pages   = {3663--3674},
}

@article{Beladi2014_Acta_GBCD_Energy,
  author  = {H. Beladi and N. T. Nuhfer and G. S. Rohrer and X. Liu and K. F. Russell},
  title   = {The five-parameter grain boundary character and energy distributions of a fully austenitic high-manganese steel using three dimensional data},
  journal = {Acta Materialia},
  year    = {2014},
  volume  = {70},
  pages   = {281--289},
}

@article{GlowinskiRohrer2016_LabFrameGBNormals,
  author  = {Krzysztof G{\l}owi{\'n}ski and Gregory S. Rohrer},
  title   = {Distributions of Grain Boundary Normals in the Laboratory Reference Frame},
  journal = {Metallurgical and Materials Transactions A},
  year    = {2016},
  volume  = {47},
  number  = {6},
  pages   = {3031--3039},
}

@book{Randle2010,
  author    = {Randle, Valerie},
  title     = {The Measurement of Grain Boundary Geometry},
  publisher = {CRC Press / Taylor \& Francis},
  year      = {2010}
}

@article{Rohrer2010,
author = {G. S. Rohrer and J. Li and S. Lee and A. D. Rollett and M. Groeber and M. D. Uchic},
title ={Deriving grain boundary character distributions and relative grain boundary energies from three-dimensional EBSD data},
journal = {Materials Science and Technology},
volume = {26},
number = {6},
pages = {661-669},
year = {2010},
}

@article{Ludwig:hx5063,
author = "Ludwig, Wolfgang and Schmidt, S{\o}eren and Lauridsen, Erik Mejdal and Poulsen, Henning Friis",
title = "{X-ray diffraction contrast tomography: a novel technique for three-dimensional grain mapping of polycrystals. I. Direct beam case}",
journal = "Journal of Applied Crystallography",
year = "2008",
volume = "41",
number = "2",
pages = "302--309",
month = "Apr",
}

@book{Sutton1995,
  author    = {Sutton, A. P. and Balluffi, R. W.},
  title     = {Interfaces in Crystalline Materials},
  publisher = {Oxford University Press},
  address   = {Oxford},
  year      = {1995}
}

@online{VanderVoort2011,
  author  = {Vander Voort, George F.},
  title   = {Introduction to Stereological Principles},
  year    = {2011},
  month   = dec,
  note    = {Metallography with George Vander Voort. Accessed 2025-09-30}
}

@book{Underwood1970,
  author    = {Underwood, Ervin E.},
  title     = {Quantitative Stereology},
  publisher = {Addison–Wesley},
  address   = {Reading, MA},
  year      = {1970},
  isbn      = {0201076500},
  series    = {Addison–Wesley Series in Metallurgy and Materials},
}

@article{AdamsField1992,
  author       = {Adams, Brent L. and Field, David P.},
  title        = {Measurement and representation of grain‐boundary texture},
  journal      = {Metallurgical Transactions A},
  volume       = {23},
  pages        = {2501--2513},
  year         = {1992},
}

@Article{Hielscher2010,
  author    = {Hielscher, Ralf and Prestin, Jürgen and Vollrath, Antje},
  journal   = {Mathematical Geosciences},
  title     = {Fast Summation of Functions on the Rotation Group},
  year      = {2010},
  issn      = {1874-8953},
  month     = jun,
  number    = {7},
  pages     = {773--794},
  volume    = {42},
  publisher = {Springer Science and Business Media LLC},
}

@Article{Hielscher2013,
  author    = {Hielscher, Ralf},
  journal   = {Journal of Multivariate Analysis},
  title     = {Kernel density estimation on the rotation group and its application to crystallographic texture analysis},
  year      = {2013},
  issn      = {0047-259X},
  month     = aug,
  pages     = {119--143},
  volume    = {119},
  publisher = {Elsevier BV},
}

@article{WINTER2025120968,
title = {Quantifying and visualizing the microscopic degrees of freedom of grain boundaries in the Wigner–Seitz cell of the displacement-shift-complete lattice},
journal = {Acta Materialia},
volume = {291},
pages = {120968},
year = {2025},
issn = {1359-6454},
author = {I.S. Winter and T. Frolov},
}

@Misc{Hielscher2026,
  author    = {Hielscher, Ralf and Wünsche, Erik},
  title     = {On the Role of the Double Fourier Sphere Method in Fast Algorithms on {SO}(3)},
  year      = {2026},
  copyright = {arXiv.org perpetual, non-exclusive license},
  doi       = {10.48550/ARXIV.2602.06677},
  publisher = {arXiv},
}

@Article{Kunis2003,
  author    = {Kunis, Stefan and Potts, Daniel},
  journal   = {Journal of Computational and Applied Mathematics},
  title     = {Fast spherical {F}ourier algorithms},
  year      = {2003},
  issn      = {0377-0427},
  month     = dec,
  number    = {1},
  pages     = {75--98},
  volume    = {161},
  publisher = {Elsevier BV},
}

\section{Appendix}
\label{sec:appendix}

\subsection{Spherical Convolutions}
\label{sec:spher-conv}

The spherical convolution of an $\mathtt{ODF}$ with a spherical function
$f_{A}(\vec n_{A})$, describing some directional property with respect to the
crystal reference frame is defined as the spherical function $g(\vec n)$
with respect to the specimen
reference frame is given by the integral
\begin{equation}
  g(\vec n) = \mathtt{ODF} \ast f_{A}(\vec n) 
  = \int_{SO(3)} \mathtt{ODF}(g) \cdot f_{A}(\mathtt{inv}(g) \vec n) \d{g}.
\end{equation}
It can be interpreted as the average of the crystal property $f_{A}(\vec n_A)$ over all orientations $g$ with respect to the $\mathtt{ODF}$.

Conversely, given a spherical function $f(\vec n)$ describing some directional
property with respect to the specimen reference frame, the convolution 
\begin{equation}
  \mathtt{ODF} \circledast f(\vec n_A) 
  = \int_{SO(3)} \mathtt{ODF}(g) \cdot f(g \, \vec n_A) \d{g}
\end{equation}
averages the specimen property $f(\vec n)$ over all orientations $g$ with respect to the $\mathtt{ODF}$ resulting in a directional property with respect to the crystal reference frame. 

Assume $f$ to be given by its spherical harmonic expansion 
\begin{equation*}
   f
   = \sum_{\ell=0}^L \sum_{k=-\ell}^{\ell} \hat f(\ell,k) Y_{\ell}^{k}
\end{equation*}
and $\mathtt{ODF}$ by it expansion into normalized Wigner-D functions (generalized spherical harmonics)
\begin{equation*}
   \mathtt{ODF} 
   = \sum_{\ell=0}^L \sum_{k,k'=-\ell}^{\ell}
   \widehat{\mathtt{ODF}}(\ell,k,k') D_{\ell}^{kk'}.
\end{equation*}
Then $\mathtt{ODF} \ast f_{A}$ and $\mathtt{ODF} \circledast f$ have the spherical harmonic expansions
\begin{align*}
  \mathtt{ODF} \ast f_{A}(\vec n)
    &= \sum_{\ell=0}^L \sum_{k'=-\ell}^\ell \left( \tfrac1{\sqrt{2\ell+1}}
    \sum_{k=-\ell}^{\ell} \widehat{\mathtt{ODF}}(\ell,-k,-k') \hat f_{A}(\ell,k) \right)
    Y_{\ell}^{k'}(\vec n) \\
    \mathtt{ODF} \circledast f(\vec n_A)
    &= \sum_{\ell=0}^L \sum_{k=-\ell}^\ell \left( \tfrac1{\sqrt{2\ell+1}}
    \sum_{k'=-\ell}^{\ell} \widehat{\mathtt{ODF}}(\ell,k,k') \hat f(\ell,k') \right)
    Y_{\ell}^{k}(\vec n_A)
\end{align*}

For two rotational functions $\mathtt{ODF}$ and $\mathtt{MDF}$ the rotational convolution is defined as
\begin{equation}
  \mathtt{MDF} \ast \mathtt{ODF}(g_B) 
  = \int_{SO(3)} \mathtt{ODF}(g_A) \cdot \mathtt{MDF}(\mathtt{inv}(g_A) g_B) \d{g_A}.
\end{equation}
In terms of normalized Wigner-D functions it reads
\begin{equation*}
  \mathtt{MDF} \ast \mathtt{ODF} (g_{B})
   = \sum_{\ell=0}^L \sum_{k,k'=-\ell}^{\ell}
   \left( \tfrac1{\sqrt{2\ell+1}}\sum_{j=-\ell}^{\ell} \widehat{\mathtt{MDF}}(\ell,k,j) \widehat{\mathtt{ODF}}(\ell,j,k') \right)  D_{\ell}^{kk'}(g_{B}).
\end{equation*}

\subsection{Fast Summation of Spherical Functions}
\label{sec:fast-summ-spher}

When estimating distributions on the sphere or the orientation space, one
frequently encounters sums of the form
\begin{equation}
  \label{eq:fastsum}
  \sum_{i=1}^{K} c_{i}\,\Psi(\vec n \cdot \vec n_{i}),
\end{equation}
where $\Psi \colon [-1,1] \to \IR$ is a radial kernel function and
$\vec n_i \in \sphere^2$, $i=1,\ldots,K$, are given directions. 
A direct evaluation of \eqref{eq:fastsum} for many evaluation points $\vec n$
is computationally expensive. This cost can be reduced significantly by
exploiting the harmonic structure of $\Psi$.

To this end, we expand the kernel $\Psi$ into a Legendre series
\begin{equation*}
  \Psi(\vec n \cdot \vec m)
  = \sum_{\ell=0}^{L} \hat{\Psi}(\ell)\,P_{\ell}(\vec n \cdot \vec m),
\end{equation*}
where $P_\ell$ denotes the Legendre polynomial of degree $\ell$.
Using the spherical addition theorem, this expansion can be written in terms
of spherical harmonics as
\begin{equation*}
  P_{\ell}(\vec n \cdot \vec m)
  = \frac{4\pi}{2\ell+1}
    \sum_{k=-\ell}^{\ell}
    Y_{\ell}^{k}(\vec n)\,\overline{Y_{\ell}^{k}(\vec m)}.
\end{equation*}
Hence,
\begin{equation*}
  \Psi(\vec n \cdot \vec m)
  = \sum_{\ell=0}^{L} \hat{\Psi}(\ell)\,\frac{4\pi}{2\ell+1}
    \sum_{k=-\ell}^{\ell}
    Y_{\ell}^{k}(\vec n)\,\overline{Y_{\ell}^{k}(\vec m)}.
\end{equation*}

Substituting this expansion into \eqref{eq:fastsum} and interchanging the order
of summation yields
\begin{align*}
  \sum_{i=1}^{K} c_{i}\,\Psi(\vec n \cdot \vec n_{i})
  &= \sum_{\ell=0}^{L} \sum_{k=-\ell}^{\ell}
     \hat{\Psi}(\ell)\,\frac{4\pi}{2\ell+1}
     \left(
       \sum_{i=1}^{K} c_{i}\,\overline{Y_{\ell}^{k}(\vec n_{i})}
     \right)
     Y_{\ell}^{k}(\vec n).
\end{align*}

The inner sum corresponds to the adjoint spherical Fourier transform of the
weighted point measures $\{c_i,\vec n_i\}$, while the outer sum represents the
spherical Fourier synthesis. Both steps can be evaluated efficiently using the
non-equispaced fast spherical Fourier transform (NFSFT)
\cite{Kunis2003}.

\end{document}